\documentclass[twocolumn]{aastex701}
\usepackage{graphicx, comment}

\def\ltsima{$\; \buildrel < \over \sim \;$}
\def\simlt{\lower.5ex\hbox{\ltsima}}
\def\gtsima{$\; \buildrel > \over \sim \;$}
\def\simgt{\lower.5ex\hbox{\gtsima}}

\newcommand {\um}{$\mu$m}

\newcommand {\msun}{M$_{\odot}$}

\newcommand {\kms}{km\,s$^{-1}$}

\def\ltsima{$\; \buildrel < \over \sim \;$}
\def\simlt{\lower.5ex\hbox{\ltsima}}
\def\gtsima{$\; \buildrel > \over \sim \;$}
\def\simgt{\lower.5ex\hbox{\gtsima}}

\newcommand {\alphaco}{$\alpha_{\rm CO}$}
\newcommand {\acounits}{M$_{\odot}$/(K\,km\,s$^{-1}$\,pc$^{2}$)}

\usepackage{xcolor}

\definecolor{myred}{HTML}{BF3465}

\shorttitle{CO($3-2$) in Spiderweb SMGs}
\begin{document}

\title{Chaotic Molecular Gas in Five Dusty Star-forming Galaxies in the Spiderweb Protocluster at $z=2.16$}

\author[0000-0002-6184-9097]{Jaclyn~B.~Champagne}
\affiliation{Steward Observatory, University of Arizona, 933 N Cherry Ave, Tucson, AZ 85721, USA}
\email{jbchampagne@arizona.edu}
\author[0000-0001-7147-3575]{Helmut~Dannerbauer}
\affiliation{Instituto de Astrof\'{i}sica de Canarias (IAC), E-38205 La Laguna, Tenerife, Spain}
\affiliation{Universidad de La Laguna, Dpto. Astrof\'{i}sica, E-38206 La Laguna, Tenerife, Spain}
\email{helmut@iac.es}
\author[0000-0002-5963-6850]{Jose Manuel P\'{e}rez-Mart\'{i}nez}
\affiliation{Instituto de Astrof\'{i}sica de Canarias (IAC), E-38205 La Laguna, Tenerife, Spain}
\affiliation{Universidad de La Laguna, Dpto. Astrof\'{i}sica, E-38206 La Laguna, Tenerife, Spain}
\email{jm.perez@iac.es}
\author[0000-0002-0930-6466]{Caitlin M. Casey}
\email{cmcasey@ucsb.edu}
\affiliation{Department of Physics, University of California, Santa Barbara, Santa Barbara, CA 93106, USA}

\author[0000-0002-8412-7951]{Shuowen Jin}
\affiliation{Cosmic Dawn Center (DAWN), Denmark}
\affiliation{DTU Space, Technical University of Denmark, Elektrovej 327, DK2800 Kgs. Lyngby, Denmark}
\email{XX}
\author[]{Matthew Lehnert}
\affiliation{Universit\'{e} Lyon 1, ENS de Lyon, Centre de Recherche Astrophysique de Lyon (UMR5574), 69230 Saint-Genis-Laval, France}
\affiliation{Sorbonne Universit\'{e}, CNRS UMR 7095, Institut d'Astrophysique de Paris, 98bis bd Arago, 75014 Paris, France}
\email{XX}
\author[0000-0002-7051-1100]{Jorge A. Zavala}
\affiliation{University of Massachusetts Amherst, 710 North Pleasant Street, Amherst, MA 01003-9305, USA}
\email{XX}

\begin{abstract}

Measuring the properties of cold molecular gas available for intense star formation in galaxy protoclusters at $z>2$ is a crucial step in understanding large scale structure formation.
We present ALMA observations of CO(3$-$2) in five dusty star-forming galaxies within $\sim0.5-4$\,cMpc of the core of the Spiderweb protocluster at $z=2.16$ to measure the molecular gas mass and kinematics in the most starbursting members of the protocluster.
All five galaxies exhibit evidence for disturbed kinematics including non-Gaussian CO line profiles, irregular spatial morphology, and strong residuals when fitting the galaxies with a classical disk model.
This could be indicative of an elevated merger rate in the outskirts of the mature Spiderweb protocluster, as all of the galaxies in our sample have multiple companions detected in H$\alpha$. 
Both the gas fractions and the gas depletion timescales of the galaxies are similar to field relations at cosmic noon, indicative of the fact that their prodigious star formation rates are compensated by similarly high gas masses.
The most massive galaxies, as well as all of the galaxies identified as X-ray AGN in previous works, have gas fractions $<30$\%, compared to the sample average of 49\%, indicating declining availability of gas for star formation.
Finally, we find that the gas fractions and specific star formation rates decline with distance from the Spiderweb Galaxy, supporting the reversal of the SFR density--radius relation in high-redshift protoclusters.

\end{abstract}


\section{Introduction} \label{sec:intro}

Galaxy protoclusters \citep{Overzier2016a} -- the high-redshift progenitors to galaxy clusters -- are expected to collapse into $M_h\sim10^{14-15}$\,\msun\, structures by the time they virialize at $z\leq1$ \citep{Chiang2013a}.
These structures, typically identified as overdensities of both star-forming and quiescent galaxies, are a key population to study in order to connect galaxy evolution to the buildup of large scale structure.
Unfortunately, they are identified with myriad observing techniques of varying depth, spatial scales, and tracers at different wavelengths, so we do not have a global census that is by any definition complete.
In many cases, we are still limited to individual case studies \citep[see review in][]{Alberts2022a}.

At lower redshifts ($z\leq1.5$), fully-formed galaxy clusters with halo masses $M_h\geq10^{14}$\,$\rm M_{\odot}$ can be identified by a red sequence of massive elliptical galaxies with low star formation rates (SFR) and gas fractions \citep[e.g.,][]{Skibba2009a, Cooper2008a}.
Also, reservoirs of hot gas in the intracluster medium (ICM) are traceable through Bremsstrahlung emission in the X-ray and/or a Sunyaev-Zel'dovich (SZ) emission or absorption signature in the submillimeter regime \citep{Kravtsov2012a}.
Well before $z\sim2$, however, there is no hallmark signature of the ICM, which requires a virialized halo to shock-heat the gas to high temperatures \citep[though see][]{DZhou2025a}, and the red sequence has generally not yet emerged \citep{Overzier2016a}. 
Thus, instead of ICM tracers, protoclusters at $2<z<6$ are mostly identified as overdensities of (typically star-forming) galaxies, including those selected by narrowband imaging in Ly$\alpha$ or H$\alpha$ \citep[e.g.,][]{Venemans2007a, Hatch2011a}, H$\alpha$ or [\textsc{O III}]$\lambda$5008 emission line galaxies with slitless spectroscopy \citep[e.g.,][]{Helton2024a, Champagne2025a}, Lyman-break galaxies \citep[e.g.,][]{Steidel1998a, Pudoka2024a, Jin2024a, Sillassen2024a} among others --- 
yet many of these studies focus on a single selection technique for a given structure, making it difficult to constrain potentially disparate galaxy populations across different structures.
The period of $2<z<4$, at cosmic noon, represents the ideal laboratory to explore protoclusters just before their final collapse, as this is when cosmic star formation reaches its peak \citep{Madau2014a}, plus galaxies in overdense regions are believed to experience the bulk of their star formation at $z\gtrsim2$ \citep[e.g.,][]{Collins2009a, Papovich201a}.

During this period of active stellar mass assembly, one of the most important physical questions about these dense environments is what mode of star formation one would expect to see in protocluster galaxies, e.g. whether it is characterized by intrinsically bursty star formation, merger-induced starbursts, or a smooth buildup of stellar mass over time. 
If galaxies in overdensities primarily build up their mass through short-lived bursts of star formation, consistent with rare dusty star-forming galaxies \citep*[DSFGs;][]{Casey2014a}, one may expect within overdensities a ubiquitous population of submillimeter-bright galaxies characterized by prolific thermal emission from cold dust. 
Indeed, many protoclusters spanning $2<z<6$ have shown evidence for strong overdensities of massive galaxies with intensely high star formation rates, including the Spiderweb Galaxy protocluster at $z=2.16$ \citep{Dannerbauer2014a}, SSA22 at $z=3.09$ \citep{Steidel1998a, Chapman2009a, Umehata2015a}, the Distant Red Core  at $z=4.00$ \citep{Oteo2018a, Long2020a}, and SPT2349-56 at $z=4.3$ \citep{Miller2018a, Hill2020a}.
An additional avenue to explore the conditions of star formation in such DSFGs is through observing the cold molecular gas, the necessary fuel for future star formation.
Tracing the cold ISM can be done via CO transitions \citep[e.g.,][]{Zavala2019a, Hill2020a, Champagne2021a, Chen2024a, Umehata2025a, DZhou2025a}, or, more recently, through far-infrared carbon lines like [\textsc{C I}] and [\textsc{C II}] \citep[e.g.,][]{Harrington2025a, Hughes2025a}. 
The kinematics of cold gas and gas-to-stellar-mass ratios offer key insights into whether protocluster galaxies undergo star formation driven by accretion of cold gas from the cosmic web or by mergers. 

The Spiderweb protocluster remains one of the best-studied protocluster systems at $z\sim2$, surrounding the Spiderweb radio galaxy at $z=2.16$ \citep{Kurk2000a, Miley2006a}.
It is characterized by an overdensity of LAEs \citep{Pentericci2000a} 
and a large filamentary structure around the radio galaxy revealed by narrowband H$\alpha$ imaging \citep{Koyama2013a, Shimakawa2018a} and narrowband Pa$\beta$ imaging \citep{PerezMartinez2024a, Shimakawa2024a}.
Various works have spectroscopically confirmed their redshifts to be consistent with the protocluster, either via H$\alpha$ \citep{Kurk2004a, PerezMartinez2023a} or CO \citep{Jin2021a}.
There is additionally a population of starbursting DSFGs identified by LABOCA at 870\um\, and SCUBA at 850\um\, \citep{Stevens2003a, Dannerbauer2014a, Dannerbauer2017a}.
\citet{Tozzi2022a}.
\citet{Shimakawa2023a} identify a large number of X-ray  AGN and \citet{DiMascolo2023a} reveals an SZ signature indicative of a nascent ICM.
Several studies, including the COALAS program carried out by the ATCA observatory, have endeavored to map out the cold molecular gas across the protocluster via CO($1-0$) \citep{Jin2021a, Chen2024a, PerezMartinez2025a}, finding evidence of extended molecular gas reservoirs in many of the HAEs.

In this paper we present a study of CO($3-2$) in a sample of 5 members of the Spiderweb protocluster, four originally identified as DSFGs and one as a pure HAE in \citep{Dannerbauer2014a} and later spectroscopically confirmed in CO($1-0$) \citep{Dannerbauer2017a, Jin2021a}.
Using ALMA Band 3 observations, we aim to characterize the kinematics and gas content of these galaxies, all located at the redshift of the protocluster and on the radial outskirts. 
In \S\ref{sec:data} we present the description and reduction of the ALMA data and in \S\ref{sec:anal} we detail our methods for extracting spectra and measuring molecular gas masses.
In \S\ref{sec:results} we present dynamical modeling results of each source to characterize the 3D distribution of molecular gas in the protocluster.
In \S\ref{sec:physstate} we discuss our results and contextualize the environmental dependence of the gas properties, and we conclude in \S\ref{sec:conclusions}.
Throughout this paper, to be consistent with previous Spiderweb literature, we assume a flat $\Lambda$CDM cosmology with H$_0$ = 70\,km\,s$^{-1}$\,Mpc$^{-1}$, $\Omega_{\Lambda}=0.7$, and $\Omega_{\rm M}$=0.3, in which 1\arcsec\, corresponds to 8.3 kpc.

\section{ALMA Data}\label{sec:data}
The data were taken with the Atacama Larger Millimeter Array (ALMA) in Band 3 with the C43-6 configuration.
The observations as part of program 2019.1.00244.S (PI: C. Casey) were taken on 2021 May 27.
The receivers were tuned to 94.54--98.03 GHz and 106.55--110.28 GHz, with one sideband covering the $J=3-2$ transition of CO at $z=2.16$ and the rest covering rest-frame 0.9\,mm continuum.
The total on-source integration time was 1330--1390\,s per source.

\begin{figure*}
\centering
\includegraphics[width=0.7\textwidth]{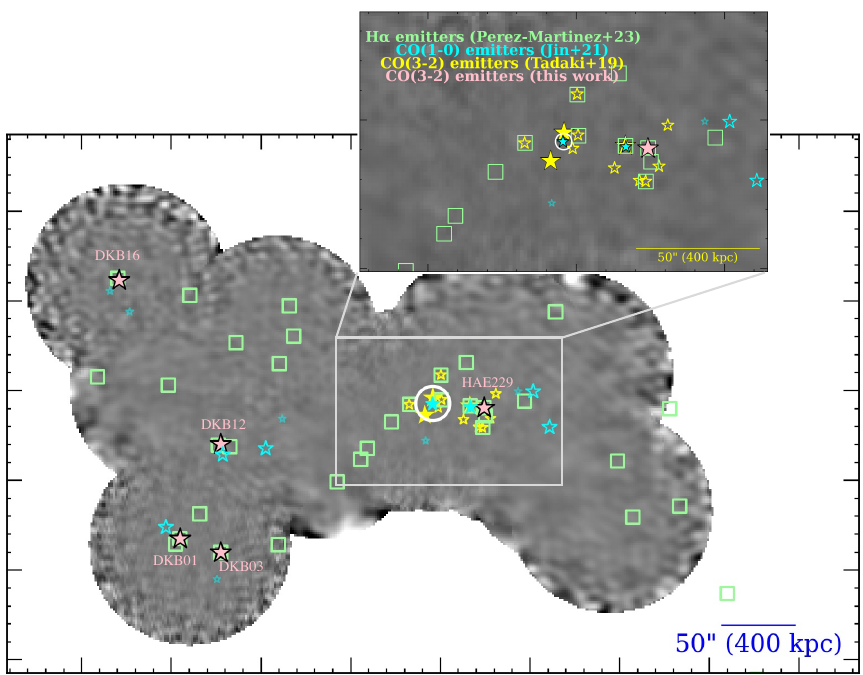}
\caption{Background: CO($1-0$) mosaic from COALAS \citep{Jin2021a} centered on the redshift of the Spiderweb Galaxy, $z=2.16$. The pink stars denote the locations of the five primary DSFGs for which we present new CO($3-2$) data from ALMA. The cyan stars mark positions of CO($1-0$) detections from \citep{Jin2021a} for which we also extract CO($3-2$) spectra, where a filled marker indicates a detection of CO($3-2$) and empty indicates non-detection. The smaller cyan stars indicate CO($1-0$) sources that could not be detected in our ALMA data due to their redshifts falling outside the tuning setup. The yellow stars mark HAEs with CO($3-2$) observations from \citet{Tadaki2019a}, where again filled (empty) indicates a (non-)detection. Other spectroscopically confirmed HAEs from \citet{PerezMartinez2023a} are shown with green squares, and the white circle indicates the Spiderweb radio galaxy. }
\label{fig:mosaic}
\end{figure*}

The sample consists of 5 galaxies originally detected at 870$\mu$m with APEX/LABOCA \citep{Dannerbauer2014a}: DKB01, DKB03, DKB12, DKB16, and HAE229 \citep[the latter also detected at 850\um\, by SCUBA;][]{Stevens2003a, Dannerbauer2014a}. 
These were spectroscopically confirmed to be members of the Spiderweb protocluster at $z\approx2.16$ through submillimeter observations of CO \citep{Emonts2013a, Jin2021a} and NIR spectroscopy covering H$\alpha$ \citep{Kurk2004a, Shimakawa2015a}.
These sources are the brightest DSFGs in the Spiderweb protocluster and are all located on its outskirts (relative to the dense core spanning $\lesssim 0.5$ Mpc), ranging from 22--176\arcsec\, (0.5--4.6 pMpc) from the Spiderweb Galaxy.

All of these sources have previous observations of CO($1-0$) as part of the ATCA project COALAS \citep{Jin2021a}, which we re-analyze here for consistent comparison of the different CO transitions.
There are four ALMA pointings centered on the locations constrained by LABOCA, except in the case of DKB01 and DKB03; these two sources are closer than the size of the ALMA primary beam, so this pointing is centered on the midpoint between them.
For reference, the primary beam is approximately 60\arcsec\, across in Band 3\footnote{\url{https://almascience.nrao.edu/proposing/technical-handbook}}.
The locations of these DSFGs with respect to the Spiderweb Galaxy are displayed in Figure \ref{fig:mosaic}.

We use the Common Astronomy Software Applications \citep[CASA;][]{CASA2022a}\footnote{\url{https://casa.nrao.edu/index.shtml}} and the latest ALMA pipeline \citep{Hunter2023a} to reduce the full dataset. 
The data cubes have a channel width of 40\,\kms\, %15.07\,MHz
and a pixel scale of $0\farcs075$ per pixel ($\approx 0.6$\,kpc at $z=2.16$). 
The images were converted from visibilities via the task \textsc{tclean} but we use the dirty cubes since the sources do not have significant side lobes. 
We use natural weighting and a $uv$-taper of 0.5\arcsec\, so as not to resolve out extended emission, with a final synthesized beam size of 1.1\arcsec$\times$0.91\arcsec. 
%We trim the images to 20\% of the primary beam response. 
The typical rms in the spectra is 0.3 mJy beam$^{-1}$ for a 40 km\,s$^{-1}$ bin. 
The continuum images are produced in the same way, using the three spectral windows which do not cover CO($3-2$) at $z\approx2.16$, with a final rms of 0.2 $\mu$Jy beam$^{-1}$.
None of the sources is detected in continuum so we do not perform continuum subtraction on the cubes.

\begin{figure*}
\centering
\includegraphics[width=0.8\textwidth]{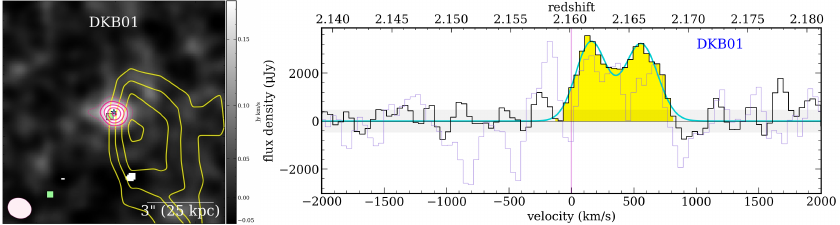}
\includegraphics[width=0.8\textwidth]{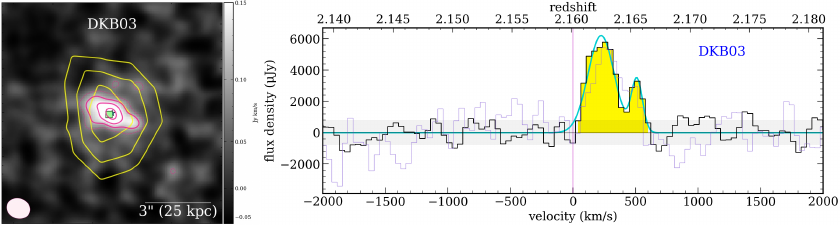}
\includegraphics[width=0.8\textwidth]{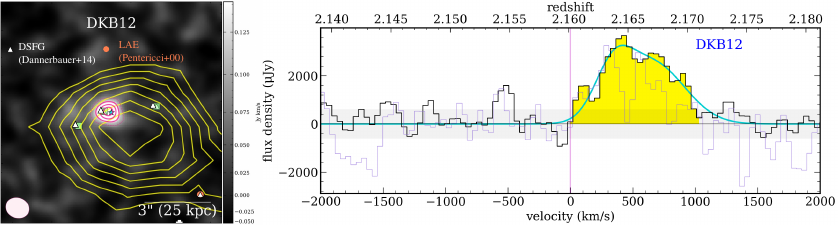}
\includegraphics[width=0.8\textwidth]{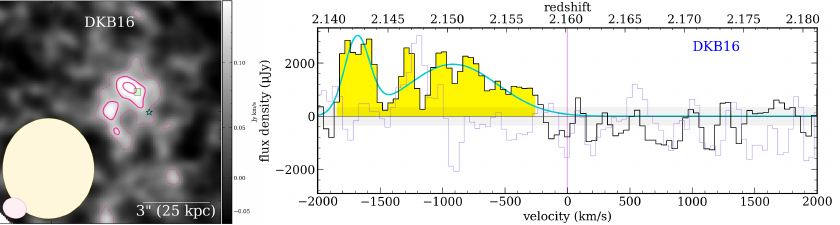}
\includegraphics[width=0.8\textwidth]{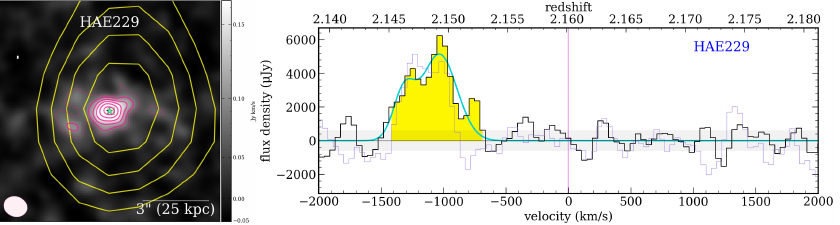}
\caption{\label{fig:spec+m0} \textit{Left:} Background: 10\arcsec$\times$10\arcsec\, moment-0 maps integrated across the channels shaded in yellow in the \textit{right} panel. The pink contours trace 3-10$\sigma$  significance in steps of 1$\sigma$ for CO($3-2$). The smoothed yellow contours show CO($1-0$) from ATCA \citep{Jin2021a} at the same significance levels. The LABOCA position is shown as a cyan star. HAEs from \citet{PerezMartinez2023a} are shown with green squares and LAEs from \citet{Pentericci2000a} in coral circles. The ALMA synthesized beam is shown as the dotted pink circle while the average ATCA beam (4.1\arcsec$\times$4.5\arcsec) is in yellow in the DKB16 panel. \textit{Right:}
ALMA CO($3-2$) spectra (black) in 40 km/s bins of all five parent DSFGs \citep[names taken from][]{Dannerbauer2014a, Dannerbauer2017a}. The average rms per channel is in shaded grey. The cyan line indicates the Gaussian fit to the spectrum, which in all cases is a multi-peaked line profile. The yellow shaded bins indicate the channels where the total flux is integrated. The thin purple line is the ATCA CO($1-0$) spectrum from the COALAS data \citep{Jin2021a}, normalized to the peak ALMA flux density and rebinned to 40 km/s channels, showing similar line profiles and centroids in all cases.
}
\end{figure*}

We also include the sources from \citet{Tadaki2019a}, who obtained CO($3-2$) measurements and upper limits in a subset of 13 HAEs closer to the core of the Spiderweb protocluster.
We do not re-reduce their data but instead take the measured CO flux values and gas masses from their paper, which we indicate where included in our analysis.
The COALAS CO($1-0$) mosaic is shown in Figure \ref{fig:mosaic}, marking the locations of the submillimeter sources and the HAEs used in this work.

\section{Data Analysis} \label{sec:anal}
\subsection{Source Extraction Procedure}

To extract spectra for each source, we use the LABOCA positions \citep{Dannerbauer2014a} and CO($1-0$) spectroscopic redshifts \citep{Jin2021a} to search for the known sources.
We obtain spectra through a custom polygonal aperture containing pixels with S/N$>$3, typically the size of 1--2 ALMA beams.
To test whether the sources were spatially resolved we also extracted a spectral profile from the brightest pixel identified in moment-0 maps (Figure \ref{fig:spec+m0}) created across 15 channels corresponding to the typical full width of each line, centered on each individual redshift.
The central pixels all agree in the shape of the line profile but are lower than the polygonal flux by factors of $1.8-2$, so we consider them marginally spatially resolved.
Figure \ref{fig:spec+m0} also shows a comparison to the CO(1-0) contours from ATCA \citep{Jin2021a}. 
All of the CO(1-0) detections are cospatial with the CO(3-2) centroids and appear marginally more extended than CO(3-2), though note the ATCA beam size ranges from (4.9--5.3)\arcsec$\times$(3.7--4.5)\arcsec. 
DKB16 is not detected in CO($1-0$); DKB12 corresponds to COALAS SW.04; and HAE229, DKB01 and DKB03 correspond to COALAS SW.01, SW.06+SW.07, and SW.03, respectively, which were identified as extended molecular gas reservoirs by \citet{Chen2024a}.

\subsection{Calculating Spectral Line Moments}\label{sec:moments}

To calculate the CO line properties we follow the procedure of \citet{Champagne2021a} (see their Appendix A), who analyzed CO($1-0$) spectra from the VLA in Hyperion, a $z=2.5$ protocluster.
We begin by fitting a single Gaussian at the expected location of each CO line, and try a double Gaussian fit if it improves the $\chi^2$.
To mitigate the effects of non-Gaussian profiles, we directly calculate the line moments and only use the Gaussian fits to determine whether galaxies contain multiple line-emitting components.
The CO redshift is determined by the first moment, or the intensity-weighted line center.
The intensity-weighted second moment $s_{\nu}$ determines the velocity dispersion, which is related to the FWHM by 2$\sqrt{2 \rm ln 2} s_{\nu}$.
The moments are recalculated 2000 times with a Monte Carlo resampling of the rms in the spectrum.
The peak flux, center, and FWHM are determined by the medians of the final distributions of the 2000 realizations, and the errors as the standard deviations of these distributions.
The integrated line intensity $I_{\rm CO}$ is the sum of the flux within the range of channels where the signal remains positive between $\pm2\sigma$ of the line center.

The spectra of all 5 sources and their Gaussian fits, extracted from the polygonal apertures, are shown in Figure \ref{fig:spec+m0}; their properties can be found in Table \ref{table:co}.
It is immediately clear they all show complex line profiles, justifying the use of line moments over integrating Gaussian fits.
All of them show at least two Gaussian components of similar widths, which could either indicate ordered rotation or merging components along the line of sight.
In particular, DKB16 is marked by several distinct components which require more than two Gaussians to describe the emission, though all of them have low SNR. 
Note that the LABOCA beam size is 19\arcsec\, so many of the sources from \citet{Dannerbauer2014a} have multiple H$\alpha$-emitting counterparts with close spatial and redshift separations (within 5\arcsec\, on sky and $\Delta z < 0.01$).
This supports the idea that many of the sources in this sample are gas-rich mergers, though see Section \ref{sec:individ} for details on individual sources.

\begin{figure*}
\centering
\gridline{
\includegraphics[width=1.0\columnwidth]{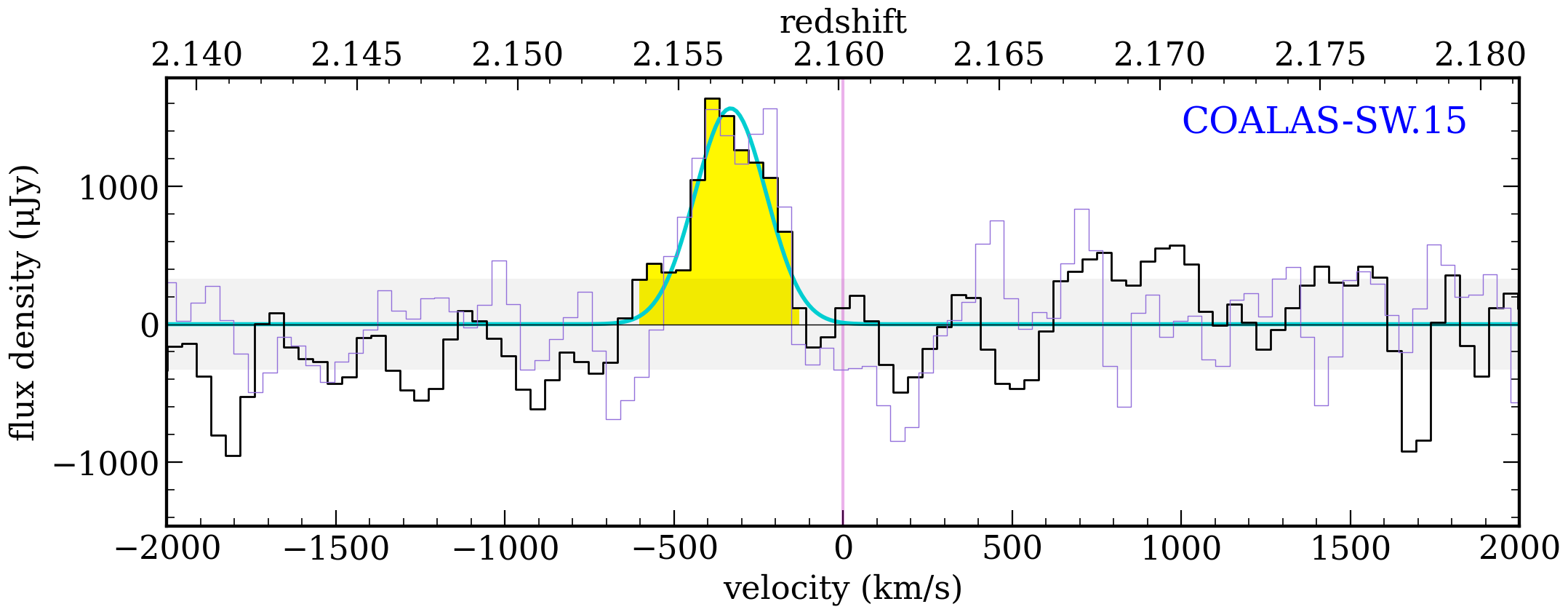}
\includegraphics[width=1.0\columnwidth]{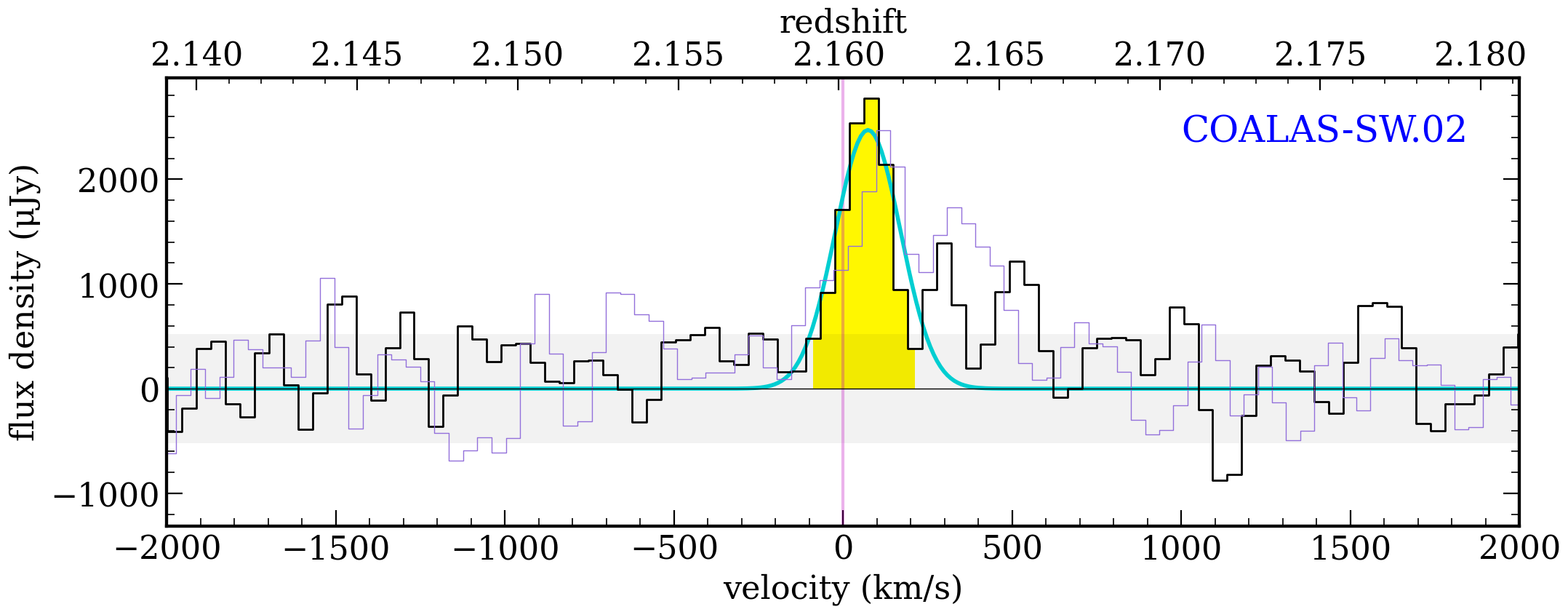}
}
\gridline{
\includegraphics[width=0.5\columnwidth]{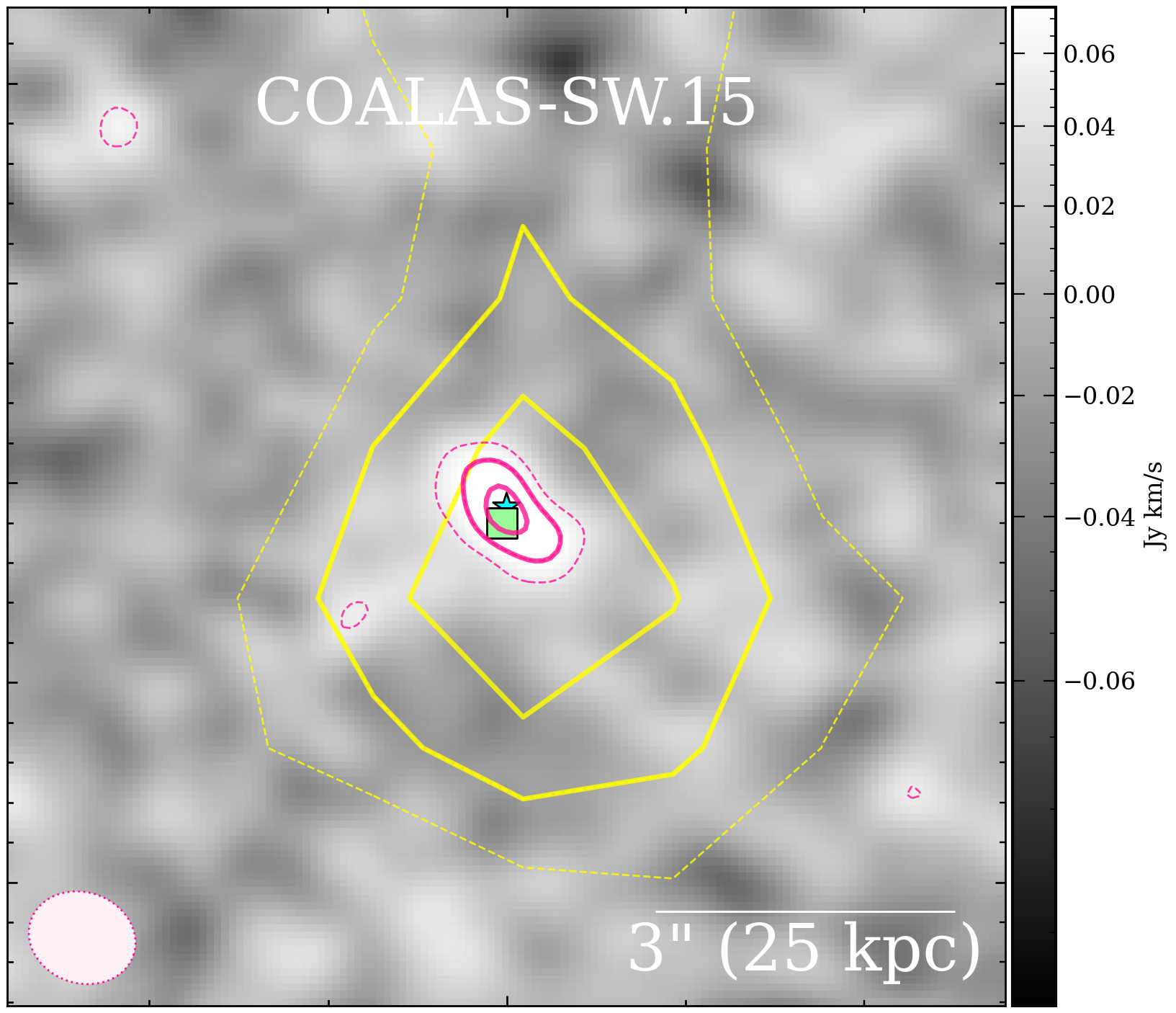}
\includegraphics[width=0.5\columnwidth]{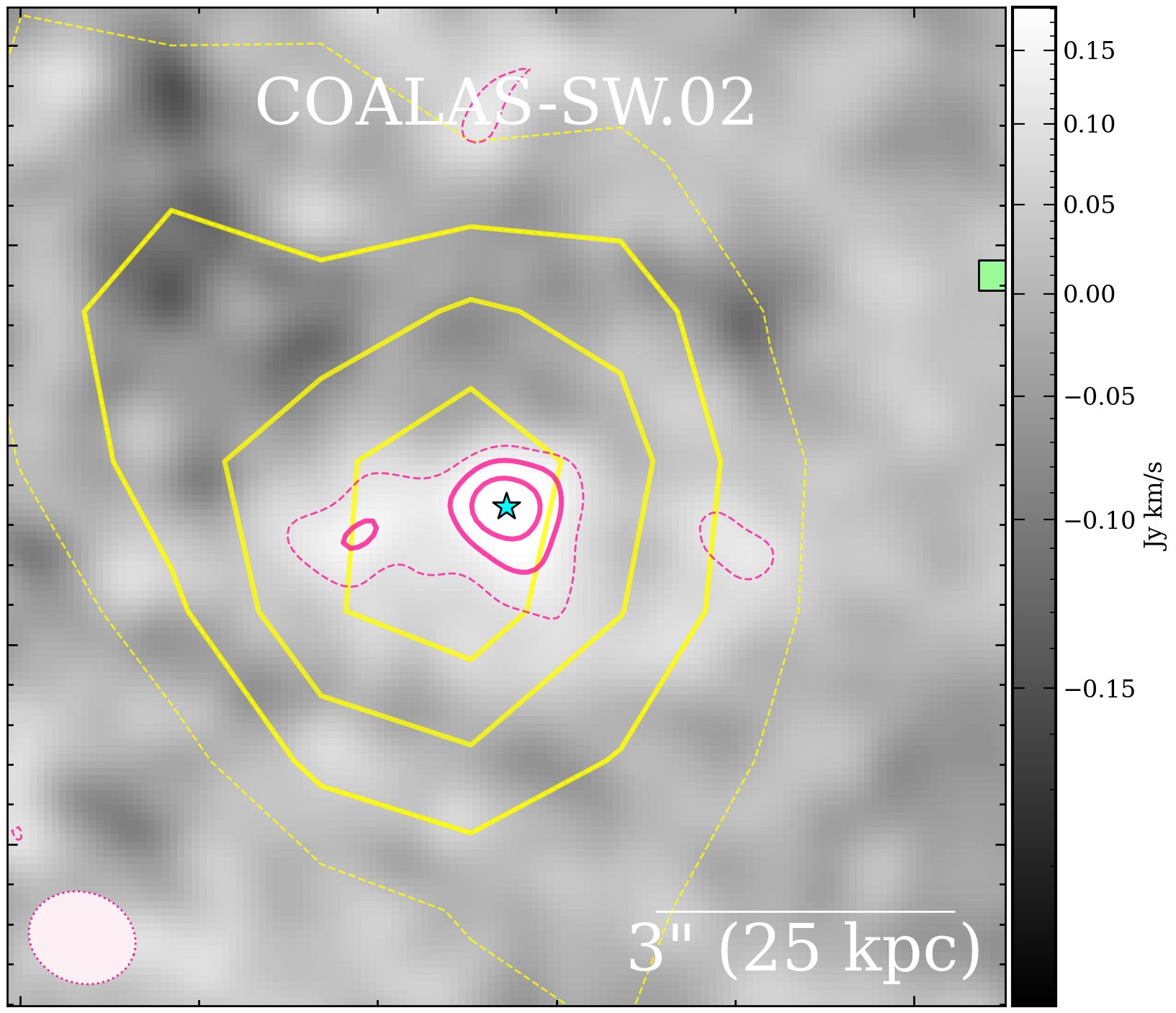}
}
\caption{\label{fig:spectra2} ALMA CO($3-2$) (black) and ATCA CO($1-0$) spectra (purple) and moment-0 maps of the two COALAS \citep{Jin2021a} detected in our data: COALAS-SW.15 and COALAS-SW.02 (the Spiderweb Galaxy). Plot symbols are the same as in Figure \ref{fig:spec+m0}. The dashed contours denote 3$\sigma$ and the solid contours increase in units of 1$\sigma$. The Spiderweb Galaxy (aka COALAS-SW.02) is strongly detected in continuum so we subtract a constant value from the spectrum.}
\end{figure*}

We also searched the datacubes for additional sources, as a total of 46 CO($1-0$) sources were found by \citet{Jin2021a} within the ATCA mosaic.
We assume we will not detect CO($3-2$) sources not already discovered in the COALAS mosaic (which was  2$\times$ deeper), so we simply use the positions of the COALAS sources as priors. 
Out of 18 sources that fall within the four ALMA primary beams, one is matched to DKB03, two to separate components of DKB01, one to DKB12, and one to HAE229 (DKB16 is not detected).
One is matched to MRC1138$-$262 (the Spiderweb Galaxy) and is strongly detected in Band 3 continuum and in CO($3-2$).
Five more COALAS sources have CO($1-0$) redshifts that fall out of the frequency coverage of our ALMA tuning (COALAS-SW.14, 21, 30, 37 and 40).
For the 7 remaining COALAS sources with SNR$>$3.7 (COALAS-SW.02, 10, 15, 16, 29, 31, and 36), which were either serendipitously discovered or matched to previous HAE counterparts, we create moment-0 maps across 10 channels centered on the expected observed frequency of CO($3-2$) and extract spectra within an aperture tracing the significant moment-0 contours.
We find that one is detected in our data (SW.15); one falls into a region where the primary beam response is $<20$\% and the other five are not detected at all.
Since these five sources do not have HAE, DSFG, or JWST/NIRCam counterparts in any of the existing Spiderweb protocluster catalogs \citep{Jin2021a, PerezMartinez2023a, Zhang2025a}, we elect not to extract upper limits and assume they may have been spurious sources in the COALAS data (all had SNR$<$5 in that data).
We include the two new sources: the Spiderweb Galaxy (MRC1138$-$262/COALAS-SW.02) and COALAS-SW.15 \citep[matched to a VLT/KMOS-identified HAE by][]{PerezMartinez2023a} in our subsequent analysis, but omit the Spiderweb Galaxy from any further discussion about dynamical modeling or environmental effects.
Their spectra and moment maps are shown in Figure \ref{fig:spectra2} and line properties in Table \ref{table:co}.

\subsection{Measuring Gas Masses}\label{sec:aco}

After measuring the line intensity $I_{CO}$ from the moments procedure outlined above, the line luminosity $L^{\prime}_{\rm CO}$ is the following \citep{Solomon1997a}:

\begin{equation}
    L^{\prime}_{CO} = 3.25\times10^7\times S_{\nu}\Delta v \frac{D_L^2}{(1+z)^3 \nu_{obs}^2} \rm K\,km\,s^{-1}\,pc^2
\end{equation}

\noindent where $S_{\nu}\Delta v = I_{CO}$ is measured in Jy \kms, luminosity distance $D_L$ in Mpc, and $\nu$ in GHz (observed frame).

\begin{deluxetable*}{ccccccccc}
\label{table:co}
\tablecaption{Derived properties from the CO($3-2$) data.}
\tablehead{\colhead{Name} & \colhead{RA} & \colhead{Dec} & \colhead{$z$}  & \colhead{I$_{\rm CO}$} & \colhead{FWHM} & \colhead{r$_{31}$}& \colhead{M$_{\rm gas}$} & \colhead{$\tau_{\rm depl}^1$}\\
\colhead{} & \colhead{J2000} & \colhead{J2000} & \colhead{} & \colhead{Jy\,km\,s$^{-1}$} & \colhead{km\,s$^{-1}$} & \colhead{}  & \colhead{10$^{11}$ M$_{\odot}$} & \colhead{Myr}}
\startdata
DKB01$^{\dagger}$ & 11:40:59.606 & $-$26:30:38.904 & 2.164 & 2.05 $\pm$ 0.10 & 551 $\pm$ 30 & 1.38 & 1.62 $\pm$ 0.08 & %135 
702 \\ %updating to Jose's SFRs rather than Helmuts 2014 IR SFR
DKB03$^{\dagger}$ & 11:40:57.800 & $-$26:30:48.035 & 2.163 & 1.99 $\pm$ 0.13 & 253 $\pm$ 59 & 1.53 & 1.42 $\pm$ 0.09 & %218
751 \\
DKB12 & 11:40:57.800 & $-$26:29:35.691 & 2.166 & 2.61 $\pm$ 0.29 & 592 $\pm$ 32 & 1.31 & 2.19 $\pm$ 0.25 & %256
1604 \\
DKB16 & 11:41:02.340 & $-$26:27:46.019 & 2.148 & 2.45 $\pm$ 0.09 & 1050 $\pm$ 22 & $<$5.02 & 0.53 $\pm$ 0.02 & %64
462 \\
HAE229$^{\dagger}$ & 11:40:46.066 & $-$26:29:11.307 & 2.149 & 2.51 $\pm$ 0.12 & 446 $\pm$ 18 & 1.55 & 1.76 $\pm$ 0.08 & 344 \\ %832 \\
\hline
SW.15$^{\dagger}$ & 11:40:46.667 & $-$26:29:10.237 & 2.156 & 0.42 $\pm$ 0.05 & 278 $\pm$ 137 & 1.10 & 0.42 $\pm$ 0.05 & %456
343 \\
SW.02* & 11:40:48.304 & $-$26:29:08.617 & 2.162 & 0.49 $\pm$ 0.08 & 459 $\pm$ 69 & 4.96 & -- & --
\enddata
\tablecomments{Molecular gas mass is converted from $L'_{\rm CO(3-2)}$ via $\alpha_{\rm CO}$ = 4.36 \acounits\, and divided by $r_{31}$. $^1$Depletion timescale calculated assuming the SFRs from \citet{PerezMartinez2023a}. $^*$Corresponds to the Spiderweb Galaxy. We do not calculate a gas mass from CO($3-2$) because $r_{31}$ deviates significantly from the typical value for star-forming galaxies. $^{\dagger}$Denotes an extended CO($1-0$) reservoir \citep{Emonts2013a, Dannerbauer2017a, Chen2024a}.} \end{deluxetable*}

We use the CO($3-2$) line measurements to calculate molecular gas mass, using the usual factor \alphaco\, to convert from line luminosity to total molecular gas mass.
The appropriate value of \alphaco\, to use can be ambiguous:
previous authors have used a range from 0.8--6.5\,\acounits\, depending on the calibration to local giant molecular clouds or luminous infrared galaxies \citep[e.g,][]{Bothwell2013a, Carilli2013a, Scoville2016a}.
Its exact value is also dependent on metallicity which may differ as a function of stellar mass \citep[see, e.g.,][]{Tadaki2019a}.
\citet{Scoville2016a} broadly argues for a higher value of \alphaco\, for use in DSFGs, especially in cases where the densest sites of star formation are not spatially resolved, as is the case here. 
However, for consistency with prior work in protoclusters at similar redshift, we use \alphaco\,=4.36\,\acounits\, which assumes a factor of 1.36 over the Galactic value for the contribution of helium and is neither dependent on stellar mass nor metallicity.
We correct all literature values used for comparison by \alphaco/4.36.

We note that the conversion of $L'_{\rm CO}$ to molecular gas mass is calibrated for CO($1-0$) and becomes progressively more uncertain for higher-order transitions when the CO spectral line energy distribution (SLED) is unknown.
To check that the ratio of $L'_{CO(3-2)}/L'_{CO(1-0)} \equiv r_{31} \sim 1$, i.e. that the gas is thermalized \citep[e.g.,][]{Kirkpatrick2019a}, we re-extract CO($1-0$) spectra from the ATCA data.
We extract spectra within apertures the size of the synthesized beam (4.5\arcsec\, on average).
Shown in Figure \ref{fig:spec+m0}, we find the same line profile shapes and similar surface brightness for each source, finding an average $r_{31} = 1.48 \pm 0.27$, which is typical for star-forming galaxies \citep[e.g.,][]{Genzel2012a}. 
Thus, we calculate the molecular gas mass from the higher-SNR CO($3-2$) but divide by $r_{31}$.
%More in-depth analysis of the CO SLED in Spiderweb will be presented in Du et al. (in prep.).

\subsection{Dynamical Modeling}\label{sec:dynmod}
The main advantage of our new CO($3-2$) data is the much higher spatial resolution than the CO($1-0$) data from \citet{Jin2021a}, whose average mosaic beam size was approximately 4.5\arcsec\, across.
Because our sources are marginally spatially resolved, we attempt to perform dynamical modeling of each DSFG to evaluate the kinematic states of each galaxy.
For this exercise, we wish to have the highest spatial resolution possible.
Since we do not see evidence for extended emission in the $uv$-tapered data cubes, we re-image each cube with no $uv$-taper and achieve a synthesized beam size of 0.33\arcsec$\times$0.45\arcsec.

To determine whether the DSFGs in our sample show signs of ordered rotation or if their line profiles are suggestive of mergers,
we attempted to fit tilted-ring models to the non--$uv$-tapered images using {\tt 3DBarolo} \citep{DiTeodoro2015a}.
We followed the methodology of \citet{Jones2021a} who used {\tt 3DBarolo} to fit the profiles of ALPINE galaxies at $z\sim4.5$ observed with ALMA.
Namely, the maximum ring size for each galaxy is determined by the FWHM of the major axis from CASA's {\tt imfit} task, and the minimum radius is determined by the minor axis of the synthesized beam. 
For most galaxies, we fit 5 rings spaced by 0.07\arcsec, and discarded galaxies that could not be fit by more than one ring based on our radius parameters (DKB16 and COALAS-SW.15). 
We use the {\tt MASK SEARCH} function to search for the source in a trimmed datacube down to SNR = 3 and \verb|growthcut|=3.5, and fix the $x$ and $y$ positions according to the peak pixels in the moment 0 map.
The systemic velocity is fixed to the spectroscopic redshift derived from CO($3-2$).
The free parameters are rotational velocity {\tt VROT}, velocity dispersion {\tt VDISP}, and position angle {\tt PA}. 

We will discuss the individual dynamical modeling results in the following section, though all of the galaxies we fit show strong residuals in their fitted velocity fields, hinting at the fact that none of the DSFGs in our sample are likely to be well described by a rotating disk.
We also note that some sources had insufficient SNR to derive an acceptable fit, especially DKB16 which showed multiple low-SNR kinematic components in the 1D spectrum.
All of the {\tt 3DBarolo} outputs, including the modeled flux, velocity, and velocity dispersion profiles and respective residuals are displayed in the Appendix, with an example for DKB01 shown in Figure \ref{fig:3dbarolo}.
Generally, {\tt 3DBarolo} fails to accurately capture the kinematic structure in each source. %, with strong residuals seen in the velocity and velocity dispersion models.
Our galaxies show maximum $v_{rot}/\sigma$ values between 14--49, well above the general threshold of $v_{rot}/\sigma>1$ that indicates rotation, naively suggesting that none of the galaxies is dispersion-dominated.
These values lie systematically above the general observed maximum of $v_{rot}/\sigma\sim10$ in galaxies observed at $1<z<4$ with ionized gas tracers such as H$\alpha$ and [OIII]$\lambda$5007 \citep{ForsterSchreiber2009a, Wisnioski2015a, Jones2025b}.
On the other hand, observations of molecular gas tracers at cosmic noon tend to show higher values of $v_{rot}/\sigma=10-20$ \citep{Rizzo2024a}.
Therefore, our galaxies could show evidence for rotationally-supported structures, but the uncertainty on $v_{rot}/\sigma$ is very high and we discuss in the next section why mergers might represent a more realistic scenario. 
There is also a strong caveat that, even with the higher spatial resolution data, \verb|3DBarolo| modeling is limited by data quality and SNR, as evidenced by the noisy residuals.

\section{Results}\label{sec:results}

\subsection{Comments on Individual Sources}\label{sec:individ}

Here we compare our findings with previous results on these sources. We will refer to the LABOCA catalogs from \citet{Dannerbauer2014a} as D14, the X-ray AGN catalog from \citet{Tozzi2022a} as T22, and the HAE analysis from \citet{Shimakawa2018a} as S18. The COALAS survey from \citet{Jin2021a} is J21.
Star formation rates and stellar masses are taken directly from these references.
A summary of properties and formation scenarios can be found in Table \ref{table:scenarios}.

\begin{figure*}
\centering
\includegraphics[width=0.8\textwidth]{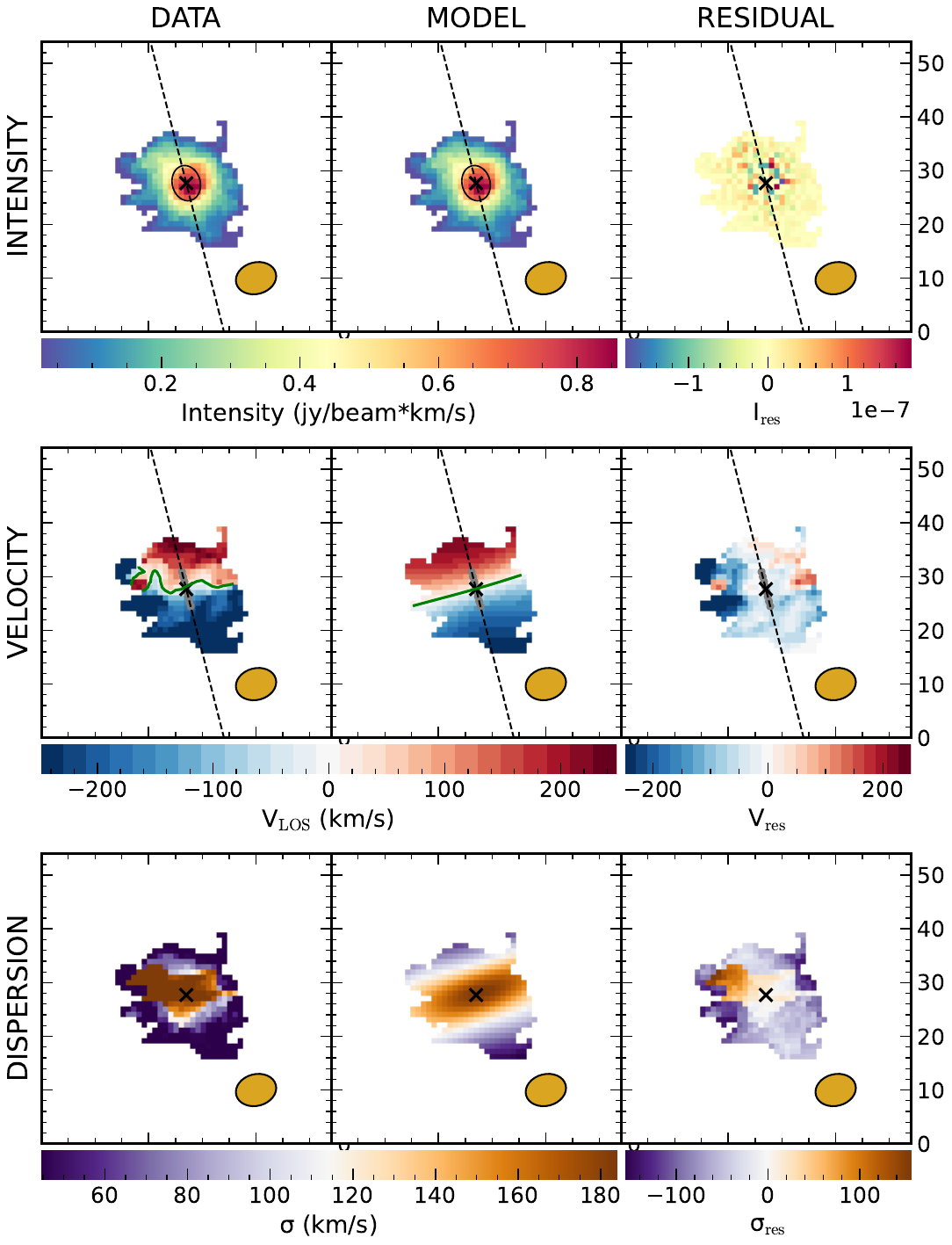}
\caption{Example fit of DKB01 using \texttt{3DBarolo}.  Each panel shows a 50$\times$50 pixel cutout around the source, with the synthesized beam shown as the gold ellipse (note that the beam is different than in Figure \ref{fig:spec+m0} since we are using non-$uv$-tapered data for this exercise). The top row shows the integrated flux (moment 0), the middle row shows the velocity field (moment 1), and the bottom row shows the velocity dispersion (moment 2). From left to right, we show the data (masked down to SNR=3), the \texttt{3DBarolo} model, and the residual. We see moderate residuals left on the edges of the galaxy in the velocity field, and the velocity dispersion is not well fit. The noisy residuals suggest that the data cannot be satisfactorily fit with a tilted ring model.}
\label{fig:3dbarolo}
\end{figure*}

\subsubsection{DKB01}\label{sec:dkb01}
DKB01 is a massive DSFG identified by LABOCA observations by D14.
Within the comparatively large LABOCA beam (FWHM 9.5\arcsec), two spectroscopically confirmed HAEs (here DKB01a and DKB01b) separated by $\sim4$\arcsec\, are associated with the source \citep[][]{Shimakawa2018a}. 
A \textit{Herschel} detection is associated with DKB01b with, at the time of that publication, $z_{phot}=2.3\pm0.7$.
In the X-ray analysis of T22, it is not associated with an X-ray AGN, and it does not appear in the low-luminosity AGN sample of \citet{Shimakawa2023a}.
J21 detected two components of DKB01b in CO($1-0$) with SNR=5.6 and 5.9, but no detection of DKB01a.
We also find that a non-detection of DKB01a but detect CO($3-2$) with redshift $z=2.164\pm0.001$ associated with the position of DKB01b.

\citet{Chen2024a} further identified DKB01 as a `robust' candidate for an extended CO($1-0$) reservoir.
On the other hand, we find using the CASA task \verb|imfit| that the beam-deconvolved size of DKB01 is 1.51\arcsec$\times$0.89\arcsec, indicating that it is not resolved (expected for the denser gas traced by CO($3-2$)).
Our CO($3-2$) spectrum is notably the brightest in the sample and has a broad double-Gaussian profile, with the second moment FWHM of $\approx550$\,km/s.

We show the {\tt 3DBarolo} modeling results in Figure \ref{fig:3dbarolo}, which fits well the flux profile in pixels above 3$\sigma$.
On the other hand, the velocity profile appears to have a smooth gradient and is well-described by the model in the center, but has strong residuals on the outskirts of the galaxy.
Moreover, the velocity dispersion appears centrally peaked in the moment-2 map but is poorly fit by {\tt 3DBarolo}.
All of this suggests that DKB01 cannot be described by a simple rotating disk.
We observe $v_{max}/\sigma$ = 21.6$\pm$10.6, %(453/20)$ [km\,s$^{-1}$] = 21.6, 
which is higher than the typical value of $\sim10$ observed at $z\sim2$ in molecular gas \citep{Rizzo2024a} but likely reflects the uncertainties and poor residuals in the {\tt 3DBarolo} fit.
In Figure \ref{fig:3d} we show how the double-horn line profile could indicate two merging components along the line of sight if we convert the velocity axis to a spatial axis (see \S\ref{sec:gas}).
Therefore, we conclude that it is a highly dust-obscured, gas-rich galaxy either with multiple clumpy components separated along the line of sight, or a merger of multiple H$\alpha$ galaxies.

\subsubsection{DKB03}\label{sec:dkb03}
DKB03 is a neighbor to DKB01, separated by only 25\arcsec\, (650 kpc) from DKB01 at the same redshift; therefore, it is likely they are physically associated and could merge in the future.
It is the second-brightest LABOCA source in our sample and has a \textit{Herschel} 250$\mu$m detection assumed to be associated with warm dust \citep{Dannerbauer2014a}; it is also associated to HAE 5 in S18.
It is not associated with either an X-ray detection in T22 nor does it appear in the low-luminosity AGN catalog of \citet{Shimakawa2023a}.
In the ALMA data, the CO($3-2$) line shows a distinctly asymmetric line profile with a smaller FWHM than DKB01.
It shows a complex velocity map with some evidence for a north-south gradient; the velocity dispersion is mostly constant at $\approx160$ km/s in the center of the galaxy, falling to 30 km/s on the outskirts \citep[slightly above but broadly consistent with other starbursting ULIRGs at similar redshift;][]{Perna2022a}.  
As seen in the appendix, the {\tt 3DBarolo} fit similarly shows a satisfactory fit in the flux profile but strong residuals in the velocity and dispersion models.
The fit results in $v_{rot}/\sigma\approx14.6\pm6.1$ in the outer region of the galaxy, though the high uncertainty questions the accuracy of the fit. 

\citet{Chen2024a} identified an extended CO($1-0$) reservoir for this source, among the most strongly detected in CO($1-0$).
It also has the brightest CO($3-2$) peak flux in our sample; its beam-deconvolved size is 2.06\arcsec$\times$1.46\arcsec, resolved across roughly 2 beam sizes.
Since it is associated with only one HAE, the merger interpretation is less certain for this source.
However, a gas-rich source with low H$\alpha$ flux (and thus not in any of the HAE catalogs) could still contribute to a merger.
It could also instead be subject to particularly turbulent motion associated with extreme star formation.
Gas-rich, turbulent disks can be subject to gravitational fragmentation due to violent disk instability \citep{Dekel2022a}.

\subsubsection{DKB12}\label{sec:dkb12}
This source is complex in that the LABOCA detection in \citet{Dannerbauer2014a} was resolved into four spectroscopically-confirmed HAEs between $z=2.164-2.171$ separated by $<5$\arcsec\, based on VLT observations \citep{PerezMartinez2023a} with 2 more LAEs a further $\sim5$\arcsec\, away \citep{Pentericci2000a}.
Interestingly, HAE27 from S18 is associated with an X-ray AGN: \citet{Tozzi2022a} classifies it as an extended Seyfert galaxy, while it is not included in the low-luminosity AGN catalog of \citet{Shimakawa2018a}.
We do not spatially resolve multiple CO($3-2$) sources, but we do see an extremely broad line profile that is indicative of multiple spectral components marginally resolved along the line of sight, as described by the overlapping double Gaussian fits.
In Appendix \ref{sec:appendix} we show the {\tt 3DBarolo} fit to this galaxy; we see significant residuals in the velocity and dispersion maps.
{\tt 3DBarolo} identifies significant flux on resolved scales, despite {\tt imfit} formally classifying it as not resolved (beam-deconvolved size 0.84\arcsec$\times$0.67\arcsec), the latter being consistent with \citet{Chen2024a} finding no evidence of an extended CO($1-0$) reservoir.
{\tt 3DBarolo} predicts $v_{max}/\sigma = 49\pm100$, indicating the failure of the fit.
The gas in this galaxy (or multiple galaxies) therefore does not show evidence for ordered rotation and is likely highly turbulent or clumpy.
Feedback from the central AGN or a violent disk instability could  potentially explain the disturbed kinematics in this source, but the existence of many HAE companions within $\sim40$\,kpc supports the possibility of a near-future or ongoing merger.

\subsubsection{DKB16}\label{sec:dkb16}
DKB16 was classified as a tentative LABOCA detection by D14 with SNR$\sim$4, but has been extensively associated with wide H$\alpha$ (ID 95 in S18, offset by $\approx$250 km/s from the CO redshift), spatially offset Ly$\alpha$, and a VLA 1.4 GHz detection suggestive of an AGN nature \citep[][and references therein]{Dannerbauer2014a}.
It is associated with XID 80 in \citet{Tozzi2022a} and classified by them as an obscured type II QSO, as well as a low-luminosity AGN in \citet{Shimakawa2023a} with an AGN fraction of 56\%.
Despite its high integrated flux, it has relatively low SNR in each channel in the CO($3-2$) data, which shows a chaotic line profile with up to four distinct but overlapping spectral components. 
Because of its low SNR, we are not able to successfully perform dynamical modeling on this source, and conclude that it is a highly disturbed source that could be composed of multiple components with low surface brightness gas.
{\tt imfit} formally classifies it as resolved (2.46\arcsec$\times$2.00\arcsec), but it is low-SNR and \citet{Jin2021a, Chen2024a} do not detect it in CO($1-0$) at all, indicating it is not extended.

\begin{deluxetable*}{ccccccc}
\label{table:scenarios}
\tablecaption{Comments on individual sources.}
\tablehead{\colhead{Name} & \colhead{X-ray AGN?} & \colhead{Extended gas?} & \colhead{HAE?}  & \colhead{$z_{CO}-z_{HAE}$} & \colhead{N companions} & 
\colhead{Scenario}}
\startdata
DKB01 & No & No & Yes & 0.0009 & 1 & Merger \\
DKB03 & No & Yes & Yes & 0.0035 & 0 & VDI \\
DKB12 & Yes & No & Yes & 0.0047 & 6 & Merger or VDI \\
DKB16 & Yes & Yes & Yes & $-$0.0026 & 0 & Unknown \\
HAE229 & No & No & Yes &  0.0006 & 1 &  Merger \\
SW.15 & Yes & No & Yes & 0 & 0 & Unknown \\
\enddata
\tablecomments{X-ray AGN classification comes from \citet{Tozzi2022a}. Extended CO($3-2$) is based on beam-deconvolved sizes. See text for details on companion HAEs.} 
\end{deluxetable*}

\begin{figure*}
\centering
\includegraphics[width=1.0\textwidth]{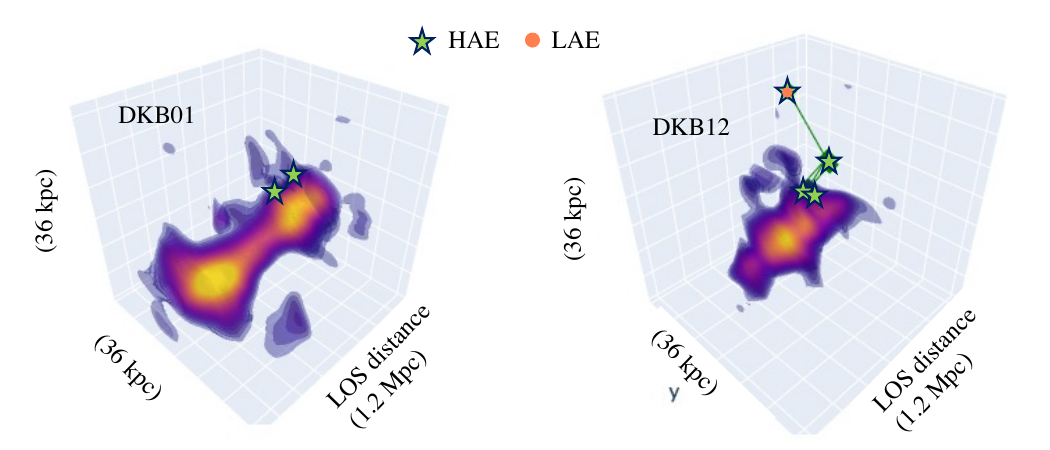}
\caption{3D visualization plots generated with \texttt{plotly}. Here we convert the frequency axis to spatial assuming that $\Delta z=0.01$ is equivalent to 1.2 Mpc, converted into a spatial direction via the comoving distance. The left panel shows DKB01 and the right shows DKB12, both with pixel values below 3$\sigma$ masked out. The surfaces show increasing significance of the emission as a fraction of the peak flux, i.e. from purple being closer to 3$\sigma$ and yellow being close to the peak flux. We can see two spatially distinct components consistent with the double-horn line flux profile in DKB01, compared to the broadened dual-component DKB12. The green stars indicate the locations of HAEs and red circles LAEs from the literature, supporting the idea that both submillimeter sources are composed of merging star-forming galaxies. Interactive plots of all sources are available online at \url{https://github.com/jbchampagne/spiderweb}.}
\label{fig:3d}
\end{figure*}

\subsubsection{HAE229}\label{sec:hae229}
``HAE229," originally selected by \citet{Stevens2003a}, has been reported to have an extended ($\sim40$ kpc) molecular gas disk traced by CO($1-0$) \citep{Dannerbauer2017a}. 
In follow-up observations, it is actually associated with two H$\alpha$ emitters separated by $<$5\arcsec\, and $\approx500$km/s \citep{PerezMartinez2023a}, with COALAS-SW.15 a further 8\arcsec\, away \citep{Jin2021a}, supporting a scenario where a merger may be ongoing or soon to occur.
It is neither associated with an X-ray detection in T22 nor a low-luminosity AGN in S23.
In CO($1-0$), J21 detected it with SNR=8.1 and \citet{Chen2024a} identified it as a robust candidate for an extended molecular gas reservoir.
In CO($3-2$) we see an overlapping two-component Gaussian suggestive of multiple components that are closely separated spatially, but {\tt imfit} formally classifies it as unresolved (0.92\arcsec$\times$0.79\arcsec).
In Appendix \ref{sec:appendix} we show the results from \texttt{3DBarolo}. 
There are significant residuals in the velocity map indicating it is not a good fit (even attempting to use the {\tt SMOOTH} parameter), and there is no evidence for a smooth velocity gradient, which could indicate distinct clumps or merging galaxies.
Much like the other sources, the high uncertainty on $v_{max}/\sigma=17.4\pm14.4$ indicates a failure to fit a rotating disk.
\citet{Zhang2025a} note that this source is morphologically consistent with a merger at short wavelengths (F182M in JWST/NIRCam) but more consistent with an extended stellar disk at rest-optical wavelengths; however, there is the possibility that the bluer filters resolve the source into multiple stellar clumps rather than a merger.
However, this does not contradict our finding in molecular gas that the HAE pair may be in the process of merging.

\subsubsection{COALAS-SW.15}\label{sec:coalas15}
COALAS-SW.15 was not part of the LABOCA sample but was extracted by J21 based on a positional HAE prior (ID 48 in S18) with SNR=5.1; \citet{Chen2024a} indicates an extended gas reservoir.
It is associated with an X-ray AGN in T22 and a low-L AGN with f$_{\rm agn}$ = 0.16 in \citet{Shimakawa2023a}.
It has a relatively faint and compact CO($3-2$) detection compared to the other sources in our sample and as such has the smallest gas mass.
Its low SNR prevents more detailed dynamical modeling or a double-Gaussian fit, so we simply declare this as a single-component HAE with detectable molecular gas indicating ongoing star formation.

\subsection{Gas in Three Dimensions}\label{sec:gas}
We showed previously that \verb|3DBarolo| cannot accurately model the kinematics of our galaxies, though we are limited by spatial resolution and SNR.
Assuming the physical failure of standard tilted-ring models to describe the morphology and kinematics of the DSFGs in our sample, we consider whether their 3D gas distributions support the idea that they are mergers.
\citet{Champagne2021a} suggested two scenarios for the configuration of molecular gas in seven closely separated galaxies in the $z=2.51$ Hyperion protocluster core. 
One is that the offsets in velocity space (i.e. redshift) of individual CO($1-0$) detections from the centroid of the protocluster core corresponded to physically extended filaments of gas along the line of sight.
The other scenario was that the velocity offset corresponded to ordered rotation of circumgalactic gas around a common gravitational center, i.e. a virialized system with a measurable velocity dispersion.
Given that the DSFGs in our sample are $>1$ Mpc from the Spiderweb Galaxy and thus are likely not yet gravitationally bound to the forming cluster \citep{PerezMartinez2023a}, we hypothesize that the gas extended along the line of sight reflects physical depth rather than virial motion within a gravitationally bound system.
This is consistent with the fact that \citet{Chen2024a} identifies extended CO($1-0$) reservoirs in half of our sample, and that the line profiles of CO($1-0$) and CO($3-2$) are very similar (Figure \ref{fig:spec+m0}), such that the denser gas traced by CO($3-2$) has a similar extent to CO($1-0$).

Here we evaluate whether the multi-component DSFGs can be described by extended filaments or mergers on the unbound outskirts of the protocluster by creating a 3D visualization converting the frequency direction into a spatial offset, in contrast to the \verb|3DBarolo| modeling.
We do this by taking the difference in comoving distance between two 40 km/s channels to spatial ``pixels," which is roughly 60 kpc.
We show two examples of this 3D visualization in Figure \ref{fig:3d} (the rest are available online): DKB12, which shows an apparently extended filament along the line of sight, and DKB01, which shows two spatially distinct, interacting components of gas.
DKB01, originally a single LABOCA source, was shown to resolve into two HAEs by \citet{Shimakawa2018a} with $\Delta z=0.0025$, as indicated in Figure \ref{fig:3d}.
The double-horn profile seen in the spectrum clearly appears as two separate components along the line of sight connected by a bridge -- cospatial with the HAEs -- which could indicate a merger.
Similarly, DKB12 resolves into 3 HAEs separated by $<1$\arcsec\, on the sky and maximum $\Delta z=0.0013$.
Two of these directly overlap with the gas reservoir, which is irregular in shape both along the line of sight and on the sky, potentially indicating a later stage merger than DKB01.
There is another nearby LAE as shown in the figure, which could be part of the system.
In both sources, the HAEs are spatially coincident with the gas reservoirs, though the molecular gas extends beyond the HAEs.
The gas is clearly further extended along the line of sight in both sources, perhaps reflecting the complex kinematics of gas in merging galaxies.
Therefore, the failure of the dynamical modeling could reflect the fact that these galaxies are not experiencing ordered rotation of the cold molecular gas but instead are the result of merger-induced starbursts.

\section{Discussion}\label{sec:physstate}

\subsection{Gas Properties}\label{sec:props}
\begin{figure*}
\centering
\includegraphics[width=1.2\columnwidth]{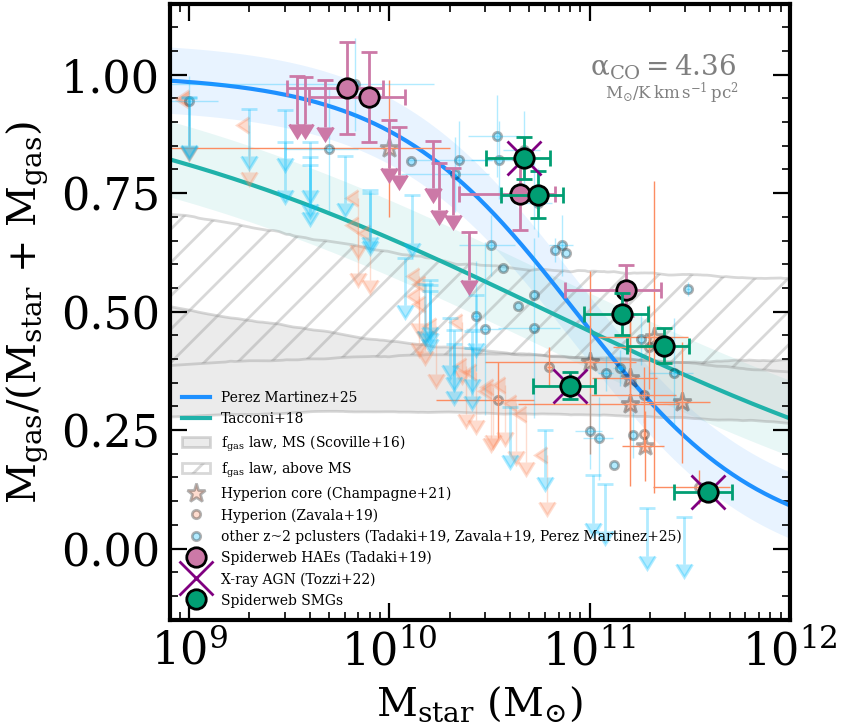}
\caption{\label{fig:fgas} Gas fraction as a function of stellar mass. For comparison we show the field scaling relation from \citet{Scoville2016a} in grey, which shows main sequence galaxies in the filled region and starburst galaxies in the hatched region. Member galaxies of the Hyperion protocluster at $z=2.49$ are shown in peach circles \citep{Zavala2019a} while the Hyperion core at $z=2.51$ \citep{Champagne2021a} is shown as peach stars. A sample of other protoclusters at $z\sim2$ from \citet{Tadaki2019a, Zavala2019a} is shown in cyan (arrows indicate upper limits). HAEs that are members of the Spiderweb protocluster with values measured from CO in previous studies are shown in pink \citep{Tadaki2019a} and the DSFGs from this work are shown in green. Those designated as X-ray AGN are shown with purple $\times$'s \citep{Tozzi2022a}. The logistic gas fraction versus stellar mass relation fit by \citet{PerezMartinez2025a} is shown in blue and a model for main sequence galaxies by \citet{Tacconi2018a} in sea green. The gas fraction in Spiderweb member galaxies is stellar-mass dependent, with lower-mass members showing an enhanced gas fraction even above starbursting galaxies, but more in line with field galaxies at very high stellar masses, consistent with the prediction of a logistic function. }
\end{figure*}

We next turn to an analysis of the gas properties in the Spiderweb DSFGs compared to the HAEs and other samples of protoclusters from the literature.
We use the total star formation rates and stellar masses derived by \citet{PerezMartinez2025a} to calculate the gas fractions and depletion timescales.

From Figure \ref{fig:fgas} we can assess the gas fraction as a function of stellar mass, in which we compare our DSFGs with other protocluster DSFGs at $z=2-3$ (all corrected to \alphaco=4.36\,\acounits).
Some of these $f_{gas}$ values, approaching 1 for lower mass galaxies, lie above even the approximately-constant ``starburst" scaling relation found by \citet{Scoville2016a}. 
However, note that \citet{Scoville2016a} evaluates total gas mass, including HI, so this is perhaps not a direct comparison.

We find instead, similar to \citet{Tadaki2019a, Zavala2019a}, that the fraction of \textit{molecular} gas with respect to stars is dependent on stellar mass, with lower-M$_*$ galaxies showing higher gas fractions. 
In contrast to the roughly flat total-gas field relation by \citet{Scoville2016a} as well as the roughly log-linear main sequence molecular gas relation of \citet{Tacconi2018a}, we find a sharp transition around log($M_*/M_{\odot}) \approx$ 10.5.
Our galaxies scatter around the best-fit logistic relation found by \citet{PerezMartinez2025a}, which was derived from the COALAS CO($1-0$) data for Spiderweb. 
Comparing with other protoclusters at $z=2-3$ \citep{Tadaki2019a, Zavala2019a, Champagne2021a}, we see that the logistic function presented by \citet{PerezMartinez2025a} accurately describes the behavior of the gas fractions of most cosmic noon protoclusters which decline with stellar mass. 
This is consistent with the predictions by \citet{Popping2012a}, who estimated molecular gas fractions for galaxies in COSMOS based on surface SFR densities.

We further note that the X-ray identified AGN \citep{Tozzi2022a} have the highest stellar masses and lowest gas fractions in our sample, suggesting that AGN feedback may be responsible for clearing out the gas in the highest-mass galaxies.
Therefore, while all of the galaxies in our sample are highly star-forming in an absolute sense (suggested by both high molecular gas masses and star formation rates, placing them on the high-mass end of the molecular gas main sequence), the gas fractions decrease with mass and AGN incidence.
This suggests that higher-mass galaxies and AGN have likely begun to deplete or expel the molecular gas available for future star formation.
This aligns well with the idea of hierarchical growth such that the highest mass galaxies have formed earlier and will quench faster than their lower-mass counterparts, consistent with other studies of molecular gas in field at similar redshifts \citep{Zhou2025a}.

\begin{figure*}
\centering
\includegraphics[width=1.2\columnwidth]{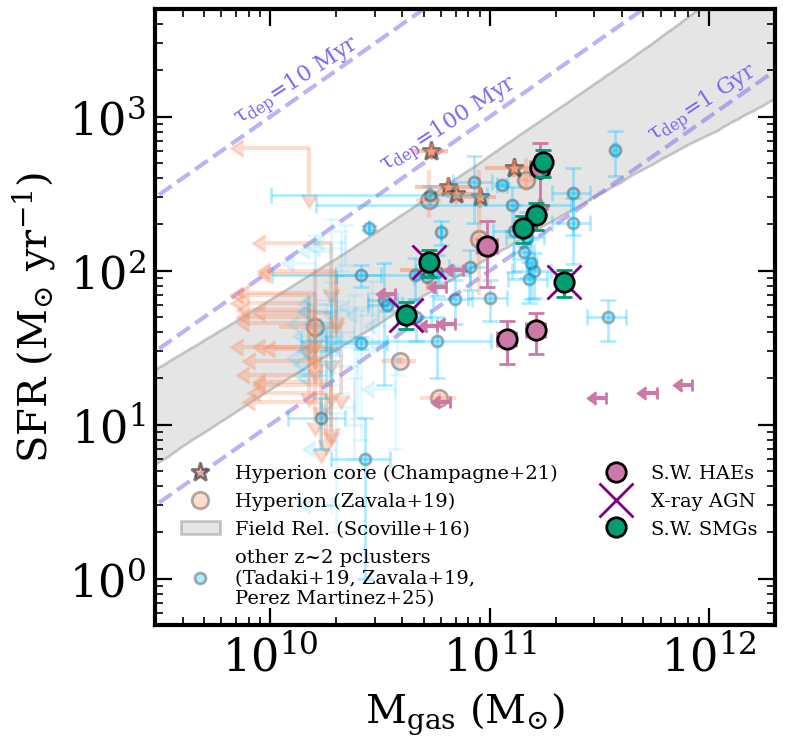}
\caption{\label{fig:sfe} Star formation efficiency, defined as SFR versus gas mass. The field scaling relation for $z=2$ galaxies is shown in filled grey \citep{Scoville2016a}. The Hyperion protocluster (circles and upper limits) and its core (stars) at $z=2.5$ are shown in peach \citep{Zavala2019a, Champagne2021a}, with other $z\sim2$ protoclusters in light blue \citep{Tadaki2019a, Zavala2019a, PerezMartinez2025a}. The DSFGs from D14 are shown in green while the HAEs from T19 are shown in pink. X-ray AGN from T22 are denoted with a purple $\times$. Lines of constant gas depletion are labeled next to the dashed purple lines. The DSFGs have gas depletion timescales on the order of $\sim$500 Myr, consistent with the field main sequence; the HAEs can instead sustain star formation for $>$1 Gyr.
}
\end{figure*}

In Figure \ref{fig:sfe} we present the star formation rate as a function of gas mass; this ratio yields the gas depletion timescale assuming constant SFR.
We note that the gas depletion timescale ($\tau$ = M$_{\rm gas}$/SFR) does not necessarily indicate a timescale for quenching: there are likely to be changes in the gas mass due to non-star-forming processes (e.g. depletion via ram pressure stripping from the ICM or AGN outflows, or conversely accretion via inflows of gas from cosmic filaments). 
However, if protocluster galaxies undergo rapid simultaneous bursts of extreme star formation induced by the overdense environment, this could be reflected by a population with very short depletion rates (Myr timescales) caused by either very high star formation rates or already-depleted molecular gas reservoirs.

The HAEs from \citet{Tadaki2019a, PerezMartinez2025a} lie below the field scaling relation with gas depletion timescales exceeding 2 Gyr, meaning they can sustain substantial stellar mass growth on long timescales if there are no interruptions to the gas reservoir.
Instead, for the DSFG population, we find shorter gas depletion timescales between 300 Myr--1.5 Gyr, occupying the same region of star formation efficiency as those in the field scaling relations of \citet{Scoville2016a}.
The DSFGs have higher stellar masses and faster $\tau_{dep}$ than the HAEs (Figures \ref{fig:fgas} and \ref{fig:sfe}), meaning that despite our proposed merger scenario, there is not enough gas supplied to sustain substantial stellar mass growth for long periods of time. 
Yet these gas depletion times are not extremely rapid compared to the field, implying that the overdense environment is not acting to substantially increase the star formation efficiency in all of the protocluster DSFGs.
Thus, similar to other protoclusters at $z=2-3$ \citep[e.g.,][]{Zavala2019a, Champagne2021a}, we do not see an elevation of star formation efficiency for members of the protocluster, implying that simply residing in an overdense environment is not the main driver of the rapid buildup of stellar mass outside the protocluster core. 

\begin{figure*}
\centering
\includegraphics[width=1.0\textwidth]{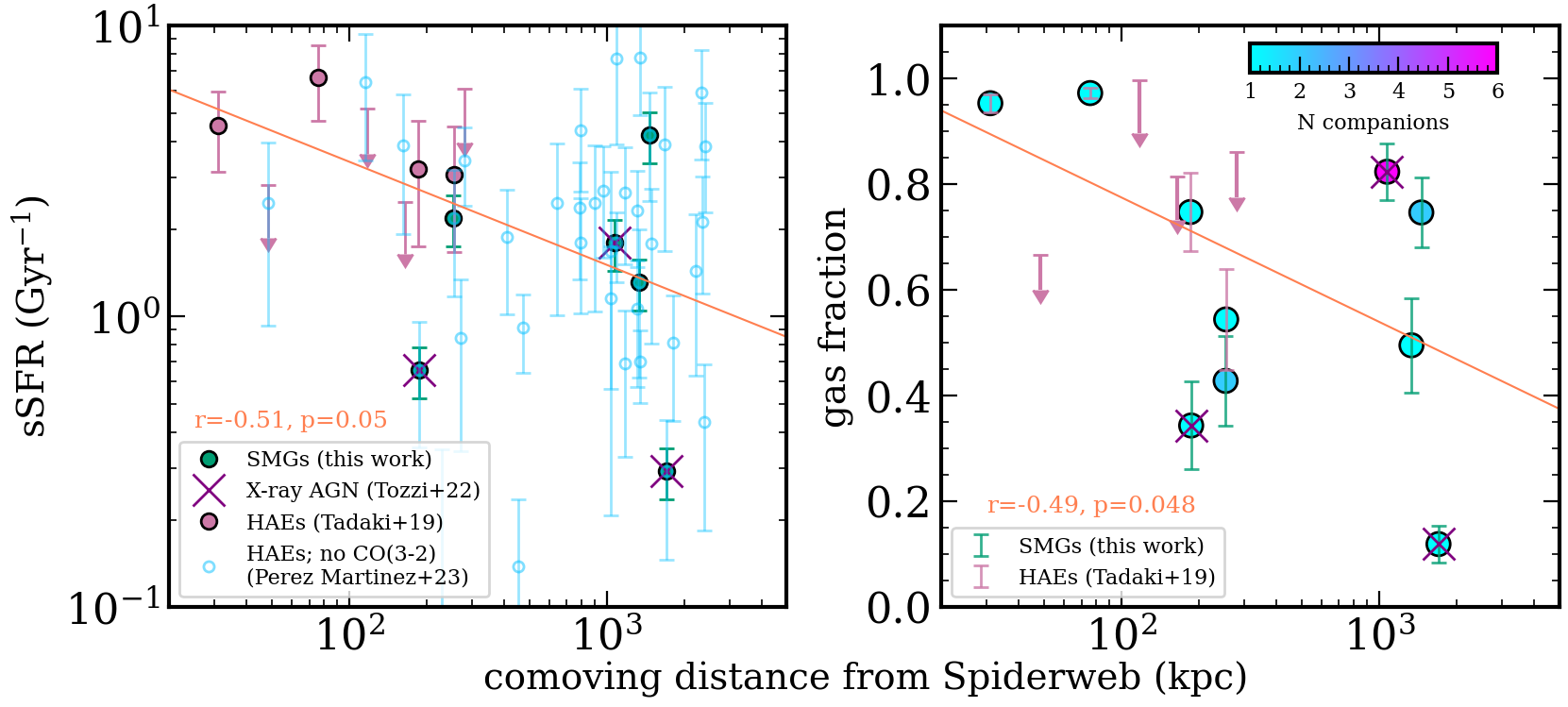}
\caption{\label{fig:radial} \textit{Left:} Specific star formation rate (sSFR) of Spiderweb protocluster galaxies as a function of transverse distance from the Spiderweb Galaxy. We display the DSFGs from this work (green; X-ray sources with purple $\times$), HAEs detected in CO($3-2$) from \citet{Tadaki2019a} (pink), and HAEs with no recorded CO($3-2$) observations from \citet{PerezMartinez2023a} (blue). \textit{Right:} Gas fraction (measured from CO($3-2$)) as a function of separation from the Spiderweb Galaxy. The colors of the points correspond to the number of HAE companions, where gas fractions are enhanced when there are $>2$ companions. In both panels, we display a simple linear regression fit in orange: both sSFR and gas fraction show tentative statistical evidence for a decrease with larger distances from the core of the protocluster.}
\end{figure*}

Generally, protocluster studies at high redshifts ($2<z<5$) have found that the DSFGs in protocluster cores gas depletion timescales on the order of 500 Myr. 
For example, \citet{Long2020a} found depletion times of $\sim$300\,Myr for red \textit{Herschel}-selected galaxies in the DRC protocluster core at $z=4.002$, expected for the field at that redshift. 
\citet{Miller2018a} also found a depletion timescale of $\sim$500\,Myr (correcting for \alphaco) in the extremely IR-luminous SPT2349-56 core at $z=4.3$. 
\citet{Champagne2021a} and \citet{Zavala2019a} found depletion timescales on the order of hundreds of Myr for gas-rich members of the Hyperion protocluster at $z=2.5$ which showed no difference from field scaling relations.
In contrast, \citet{Emonts2016a} found a longer depletion timescale of 1.2\,Gyr for the central radio galaxy in the Spiderweb protocluster (MRC1138).
For non-DSFGs, \citet{PerezMartinez2025a} find even longer $\tau_{dep}=1-3$ Gyr for high-mass ($M_*>10^{10.5}$\msun) HAEs in Spiderweb with low gas fractions ($<50$\%).
The shorter depletion timescales in DSFGs could be linked to smaller gas reservoirs due to rapid consumption from starburst-like SFRs or expulsion processes such as AGN feedback.
On longer timescales ($\tau>500$\,Myr), a combination of gas over-consumption within galaxies as well as diminishing availability of cold gas from the cosmic web will eventually shut down star formation \citep*{Remus2025a}.
Taken together, depletion timescales in protoclusters at cosmic noon imply that the overdense environment does not significantly affect star formation inside galaxies (especially those that have not yet fallen into the gravitational potential well of the cluster), instead preferring in-situ star formation until the gas is exhausted or rapidly cleared by AGN feedback on hundreds Myr timescales. 

Figures \ref{fig:fgas} and \ref{fig:sfe} together suggest that lower-mass HAEs have enough gas to sustain star formation over a long period of time ($\sim$Gyr-scale) and grow into more massive galaxies, while the massive DSFGs with lower gas fractions will likely deplete their gas more quickly and with lower fractional mass growth.
If all of the molecular gas is consumed by star formation at the current rate (with no additional inflows or outflows), the depletion timescales of these protocluster galaxies imply that they can sustain their current star formation rates for 300--1500 Myr (i.e., until $2<z<1.6$), at which time they will have built up stellar masses of median log\,(M$_{\star}$/M$_{\odot}$) $\approx$ 10.9--11.9.
These approach the massive end of the red sequence seen in $z\sim1$ clusters, close to maximum stellar masses seen in clusters according to the stellar mass function at $z\sim1$ \citep{vanderBurg2013a}. 
However, it is likely that as galaxies fall towards the virial center of the Spiderweb cluster, interaction with the nascent ICM \citep{DiMascolo2023a} and neighboring galaxies will strip or heat some of the molecular gas (through tidal interactions) prior to forming stars at the current SFR, on top of any secular quenching processes as discussed earlier.

\subsection{Environmental Dependence}

In Figure \ref{fig:radial} we show the specific star formation rate and gas fractions of the protocluster members as a function of transverse distance from the Spiderweb Galaxy. 
A Pearson test yields $r=-0.51, p=0.05$ for sSFR versus distance and $r=-0.49, p=0.048$ for the gas fraction, suggesting tentative evidence for a moderate inverse correlation.
The DSFGs we study in this work are on the outskirts of the protocluster relative to the core (all are $>0.5$\, cMpc from the SW galaxy) and show the lowest specific star formation rates and a range of gas fractions.
We reiterate that the lowest gas fractions are seen in the X-ray identified AGN which implies that AGN activity is enhanced in the outskirts, which in turn creates feedback that reduces the amount of cold molecular gas available for star formation  (reflected by the lower sSFRs).
The HAEs show a similarly negative correlation between sSFR and distance, whereby star formation remains intense and molecular gas abundant in the inner regions of the protocluster.

While the absolute values of sSFR and gas fraction in the DSFGs in our sample remain high, i.e. these galaxies are still experiencing rapid stellar mass assembly, there is a clear transition at higher stellar masses where incidence of AGN increases and gas fraction declines.
This suggests that inflowing gas is not enough to compensate the current gas consumption plus heating due to AGN feedback or shocks, which in turn prevents cold gas from forming more stars. 
The environment may contribute to diminish the efficiency of such inflows, while interactions with nearby galaxies may provide initial fuel for starburst activity (indicated by enhanced gas fractions in galaxies with more companions).
However, internal secular processes like AGN feedback and gas consumption -- rather than environmental processing -- will eventually shut down star formation in these DSFGs.
This is consistent with the fact that most of the quenched galaxies in Spiderweb are instead found within $R_{200}$ \citep{Naufal2024a}, where rapid processes like ram pressure stripping and longer processes like gas strangulation are more efficient \citep{Boselli2022a}. 
In summary, we find that the specific star formation rate and gas fraction both decline as a function of distance from the Spiderweb galaxy, but because we find depletion timescales and star formation efficiency consistent with field scaling relations, we interpret this as evidence of secular processes and mergers primarily driving the gas properties rather than global environmental processes.
With high absolute values of the gas fraction \citep[the lowest gas fraction we find is $\approx$10\%, in contrast to CO-undetected, gas-poor galaxies in lower-redshift clusters, e.g.][]{Rudnick2017a}, active star formation will be sustained for several hundred Myr, as evidenced by spectro-morphological evidence for mergers and extended gas reservoirs.

\begin{figure}
\centering
\includegraphics[width=1.0\columnwidth]{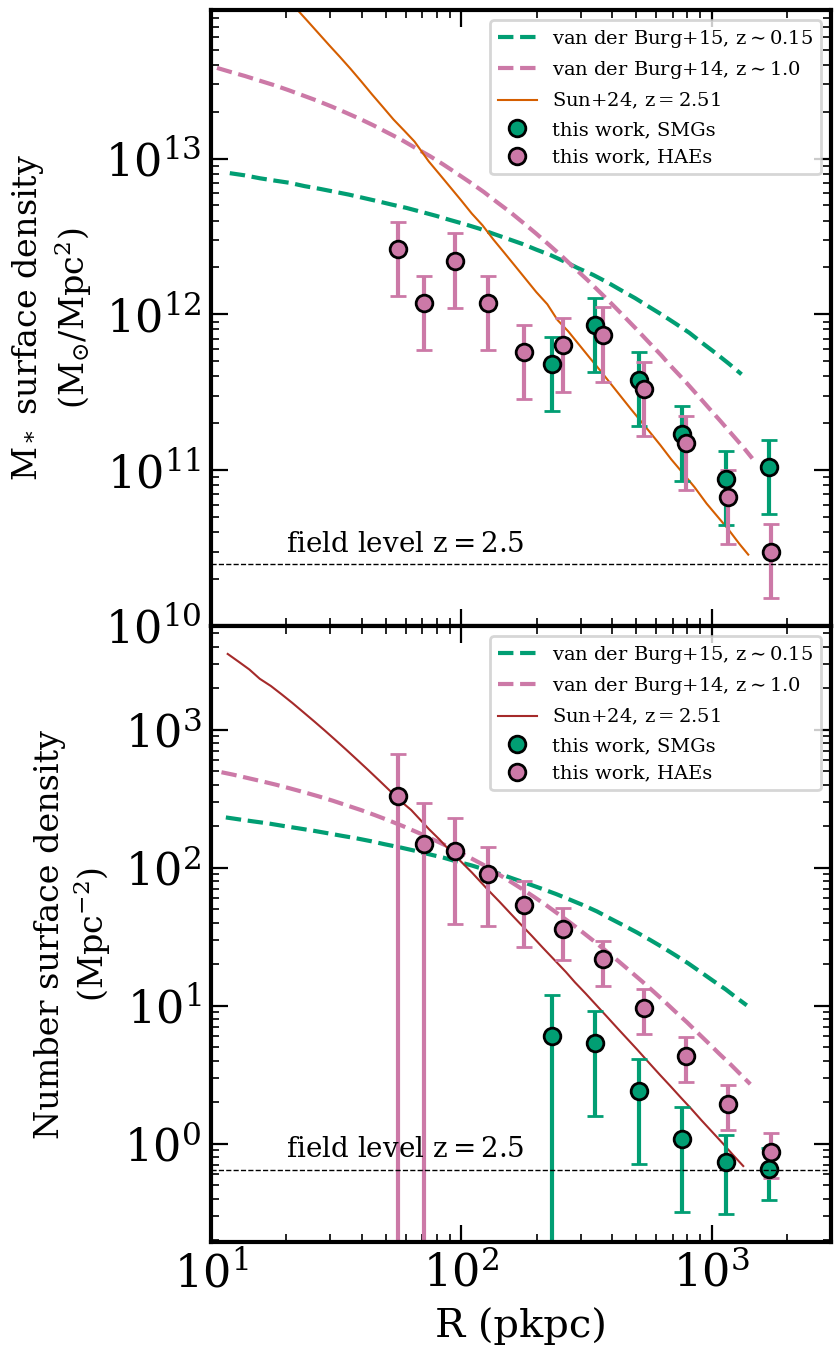}
\caption{\label{fig:density} \textit{Top:} The stellar mass surface density profile of DSFGs and HAEs in the Spiderweb protocluster that have log(M$_*$/M$_{\odot}$)$>$10 and measurements of CO(3-2). For reference, we show the generalized NFW profiles for clusters at $z\sim1$ and $z\sim0.15$ from \citet{vanderBurg2014a, vanderBurg2015a}, converted to physical units assuming $M_{200} = 3\times10^{14}$\,\msun\, at $z=1$ and $M_{200} = 9\times10^{14}$\,\msun\, at $z=0.15$. We also show the fit NFW profile to the Hyperion protocluster core at $z=2.5$ from \citet{Sun2024a}. The field level is taken from the COSMOS2020 catalog \citep{Weaver2023a}, evaluated at $z=2.5$. \textit{Bottom:} Similar to the \textit{top} plot, but displaying the surface number density of HAEs and DSFGs in Spiderweb, along with the same profiles from the literature. The steep concentration of HAEs in the center is similar to Hyperion and contrasts with the flatter profiles seen in virialized clusters at $z\leq1$. }
\end{figure}

Finally, in Figure \ref{fig:density} we show radial surface density profiles for the galaxy number and stellar mass in this sample, including both DSFGs and HAEs that have measurements of CO($3-2$).
All of these galaxies have log(M$_*$/M$_{\odot}$) $\gtrsim 10$ for fair comparison to other samples at lower and higher redshift.
The solid line in both panels compares our observed profile to that of CLJ1001/Hyperion at $z=2.5$ \citep{Sun2024a}, who reported a very high concentration of galaxies in the core of the protocluster, with a much steeper slope than is observed in virialized clusters at $z<1$.
Indeed, we find qualitatively that our sample of HAEs and DSFGs closely follow the stellar mass density profile of CLJ1001 at $R>100$ proper kpc, while the HAE profile flattens at closer separations.
In contrast, our sample at $z=2.16$ matches more closely to the number density profiles derived by \citet{vanderBurg2014a} for a sample of clusters at $z\sim1$, with $c=R_{200}/r_s =$ 2.45 for the star-forming population.
We do not fit an NFW profile to our sample simply because the full extent of the protocluster including the outskirts DSFGs is beyond the typical virial radius at lower redshift and we are incomplete to other member galaxies without CO($3-2$) observations.
Therefore the DSFGs are consistent with in-situ formation away from the core. %, and we are also incomplete to other member galaxies missing CO observations. 
However, we show qualitatively that this protocluster is consistent with the high concentration values found for high redshift clusters and protoclusters.

\citet{Zhang2022a} studies two BOSS protoclusters at $z=2.2$ and find a higher overdensity of 850\,\um-detected SMGs than HAEs in both structures. 
They find a radial cluster profile in which HAEs are more strongly clustered in the center and SMGs are clustered on the outskirts. 
Similarly, \citet{Zhang2024a} find that ALMA 1.2\,mm continuum detections of Spiderweb galaxies closely align with the spatial distribution of CO($1-0$) emitters which are concentrated $>$1\,cMpc from the core; this contrasts with the more centrally concentrated HAEs which have fewer dust continuum detections.
Indeed, we find a similar distribution of HAEs versus CO-detected DSFGs in Spiderweb as shown by Figure \ref{fig:density}, where the number density of HAEs strongly peaks approximately 100 kpc from the Spiderweb Galaxy.
The high incidence of DSFGs (which are also H$\alpha$-emitting galaxies) in the outskirts of Spiderweb suggests that cold gas is readily available to inflow in galaxies on the outer regions of the protocluster, where the intracluster gas has not yet been heated \citep[i.e. cold-in-hot accretion;][]{Overzier2016a, DiMascolo2023a}.
Taken together, the high incidence of merger signatures (or disturbed kinematics), X-ray AGN activity, and concentration of DSFGs on the outskirts indicate that the outer neighborhood of the Spiderweb protocluster is in a period of chaotic formation and has not yet reached the quenched equilibrium seen in lower-redshift clusters.

\section{Conclusions}\label{sec:conclusions}
In this work, we have characterized the molecular gas in 5 spectroscopically-confirmed DSFGs on the outskirts ($>R_{200}$) of the Spiderweb protocluster at $z=2.16$.
Using ALMA Band 3 observations targeting CO($3-2$) in galaxies selected for their bright submillimeter detections, we successfully detected CO in all five galaxies, in addition to two serendipitous detections of COALAS-SW.15 and the Spiderweb Galaxy which were previously observed with ATCA in CO($1-0$) \citep{Jin2021a}.
Our chief conclusions are:

\begin{itemize}
\item All 5 DSFGs are confidently detected in CO($3-2$), with a variety of complex, multi-Gaussian line profiles ranging in FWHM from 300--1100 km/s. This indicates that the galaxies are likely composed of multiple clumps or merging galaxies along the line of sight, supported by most of the DSFGs having multiple nearby HAE companions.

\item The sources are marginally spatially resolved in the ALMA data. We attempt to perform dynamical modeling with {\tt 3DBarolo} using a tilted-ring disk model, which was formally successful for four galaxies in our sample. However, we find that while the 2D flux profiles are well fit, {\tt 3DBarolo} is unable to describe the velocity and velocity dispersion profiles with a simple disk, leaving strong residuals in the outskirts and centers of the galaxies respectively. Given that 4 out of 5 galaxies are associated to multiple HAEs within the ALMA beam size, we conclude that most of them are candidate mergers whose kinematics cannot be described by a rotating disk. This is corroborated by their irregular 3D gas distributions.

\item We observe gas masses ranging from 0.42--2.19$\times10^{11}$\,\msun\, with gas fractions (with respect to stellar mass) ranging from 0.1--0.85. Given their stellar masses of $\sim10^{11}$\,\msun\, and high IR-based SFRs taken from the literature, all of these galaxies are consistent with the massive end of the star-forming main sequence.

\item The gas fractions are elevated with respect to field galaxies at $z=2$ at fixed stellar mass, but follow a sequence of decreasing gas fraction with stellar mass consistent with the logistic mass-dependent functional form found by \citet{PerezMartinez2025a} for HAEs in the Spiderweb protocluster. The star formation efficiency increases linearly with stellar mass, similar to other protoclusters and, notably, indistinguishable from field galaxies at $z=2-3$, with gas depletion timescales ranging from 300--1500 Myr. Three galaxies are associated with X-ray detections indicating AGN activity, two of which show the lowest gas fractions in our sample. AGN feedback and gas exhaustion following mergers may be responsible for the declining gas fraction with mass.

\item While all galaxies are actively star-forming, both the specific star formation rate and the gas fraction decline with distance from the Spiderweb galaxy. This is consistent with centrally concentrated formation of the protocluster wherein galaxies close to the Spiderweb Galaxy experience accelerated stellar mass assembly. While the star formation efficiency is not found to be different from the field, the high gas masses and elevated merger and AGN activity found in our sample indicate that the overdense environment and availability of cold gas from the cosmic web induce rapid galaxy evolution on the outskirts of the protocluster as well. 

\item We measure the stellar mass and number density as a function of radial distance from Spiderweb, showing that the profiles are moderately consistent with the functional form found for the Hyperion protocluster at $z=2.5$ \citep{Sun2024a}, and lower in normalization than clusters at $z=1$ but with a similar shape \citep{vanderBurg2014a}. The HAE number density peaks closer to Spiderweb, while the gas-rich DSFGs are found on the outskirts, where gas inflow and nearby neighbors create an environment conducive to dust-obscured star formation.
\end{itemize}

None of the DSFGs shows evidence for simple rotational motion, but could instead be experiencing a compaction phase in which rapid consumption of the central gas reservoir leads to a gradual decline in the star formation, eventually resulting in a quenched galaxy \citep[e.g.,][]{Tacchella2016}.
We further posit that AGN feedback could be responsible for the lower gas fractions with respect to the field at this epoch.
Overall, we find that the Spiderweb protocluster is still assembling stellar mass at a rapid rate from the core to the outskirts, demonstrated by the chaotic gaseous environment we observe.
More observations of higher-order CO transitions in cosmic noon protoclusters will soon elucidate the main star formation mode in clusters prior to virial collapse.

%\acknowledgments

\begin{acknowledgments}
JBC thanks Bjorn Emonts for helpful discussions.
JBC  acknowledges funding from the JWST Arizona/Steward Postdoc in Early galaxies and Reionization (JASPER) Scholar contract at the University of Arizona. 
HD and JMPM acknowledge financial support from the Ministerio de Ciencia, Innovación y Universidades (MCIU/AEI) under grant (Construcción de cúmulos de galaxias en formación a través de la formación estelar oscurecida por el polvo) and the European Regional Development Fund (ERDF) with reference (PID2022-143243NB-I00/DOI:10.13039/501100011033).

\software{ CASA (CASA Team 2022), Astropy \citep{Astropy18}, Matplotlib \citep{Hunter2007a}}
\facility{ALMA, ATCA}
\end{acknowledgments}
\bibliography{main}

\clearpage
\appendix
\section{3DBarolo Fits}\label{sec:appendix}
In this Section we display the 3DBarolo fits for each DSFG. Note that the fit failed for DKB16 and COALAS.SW15 altogether due to low SNR, so we do not display plots for these two sources.

\begin{figure*}[h!]
\centering
\includegraphics[width=0.65\textwidth]{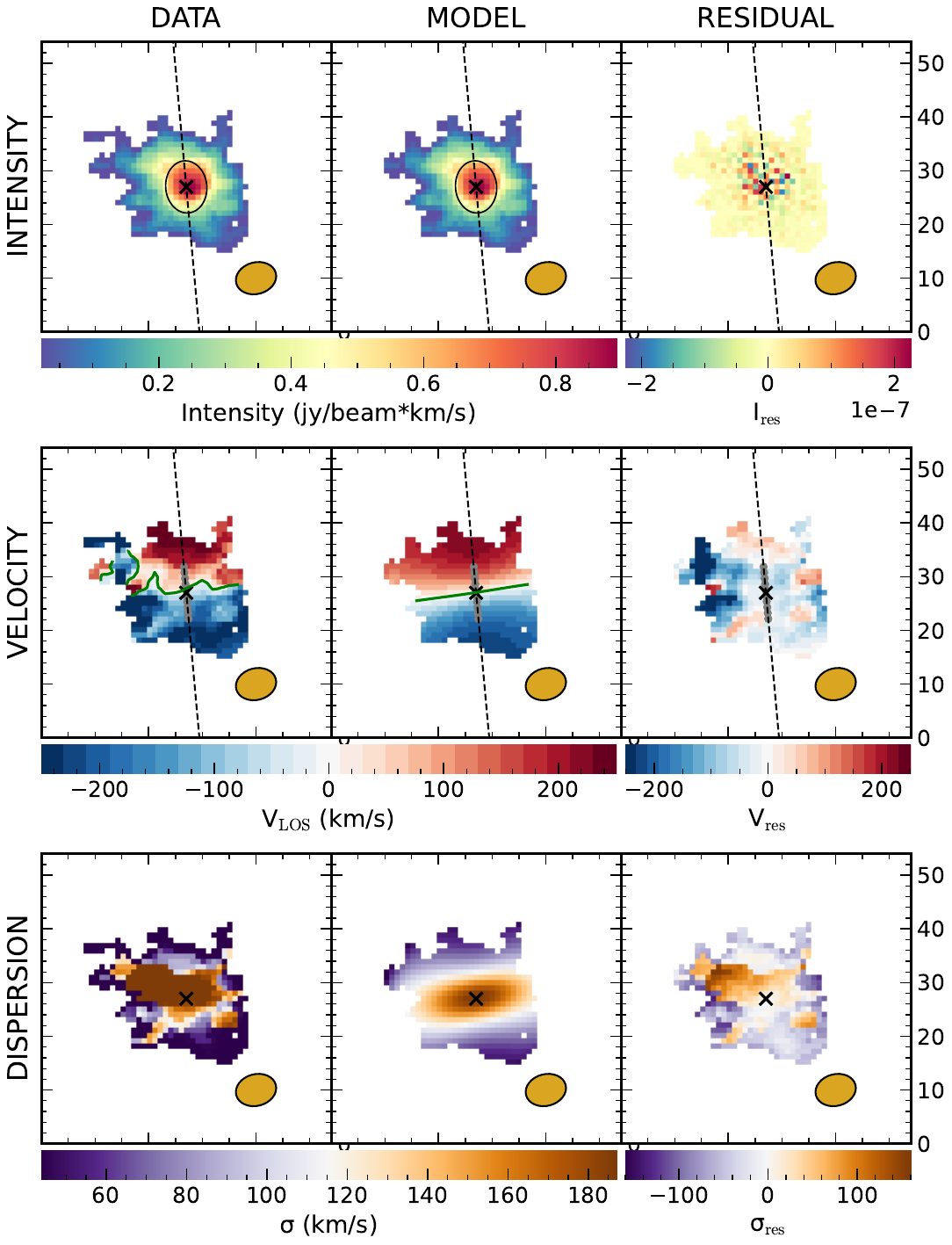}
\caption{Example fit of DKB03 using \texttt{3DBarolo}. Each panel is a 50$\times$50 pixel cutout of the non-$uv$-tapered data, with the beam size in gold in the bottom corners. The top row shows the integrated flux (moment 0), the middle row shows the velocity field (moment 1), and the bottom row shows the velocity dispersion (moment 2). From left to right, we show the data (masked down to SNR=3), the \texttt{3DBarolo} model, and the residual. There are moderate residuals left on the edges of the galaxy in the velocity field, and the velocity dispersion is not well fit in the central region of the galaxy.}
\label{fig:3dbarolo1}
\end{figure*}

\begin{figure*}
\centering
\includegraphics[width=0.65\textwidth]{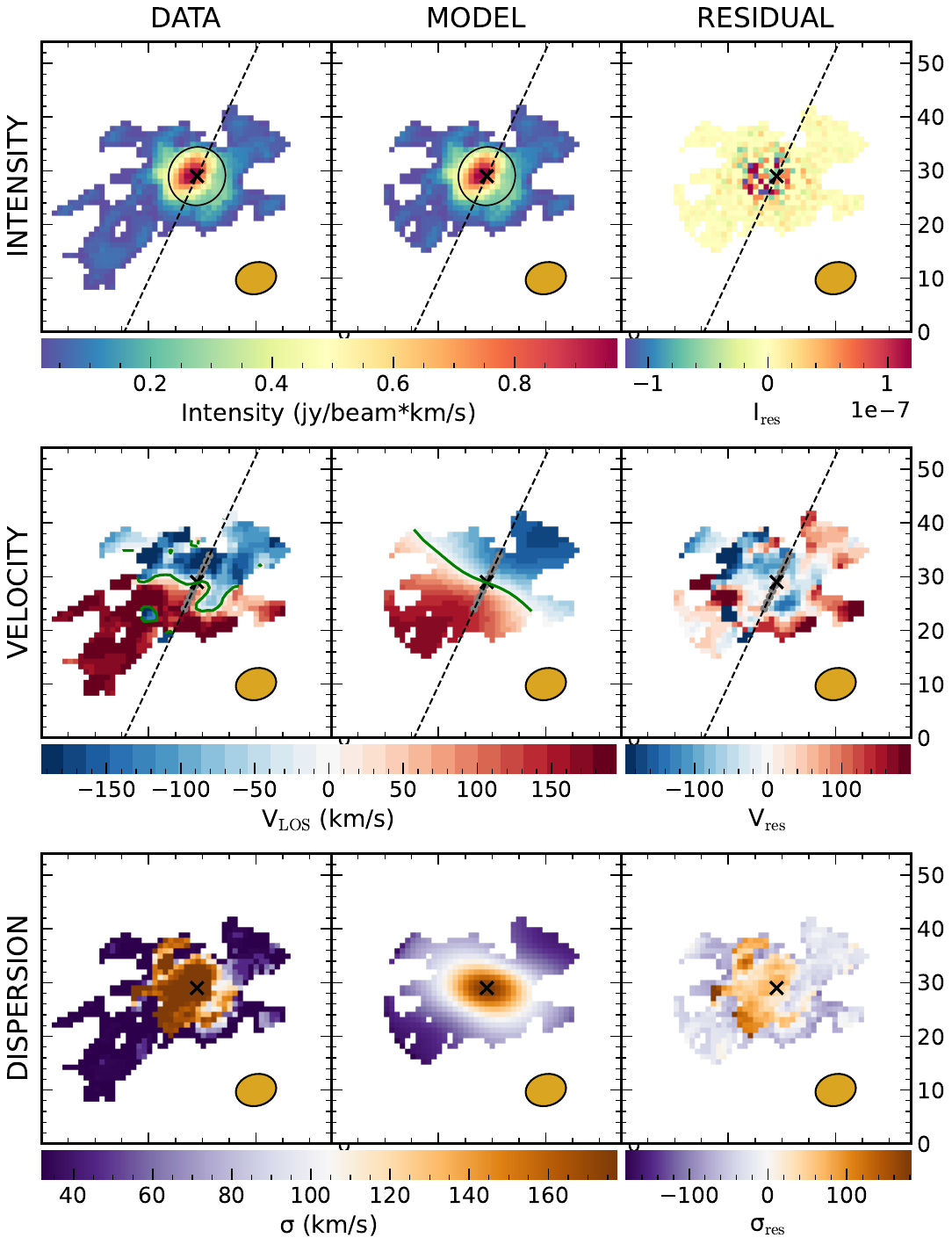}
\caption{Example fit of DKB12 using \texttt{3DBarolo}. Same as above figure.}
\label{fig:3dbarolo2}
\end{figure*}

\begin{figure*}
\centering
\includegraphics[width=0.65\textwidth]{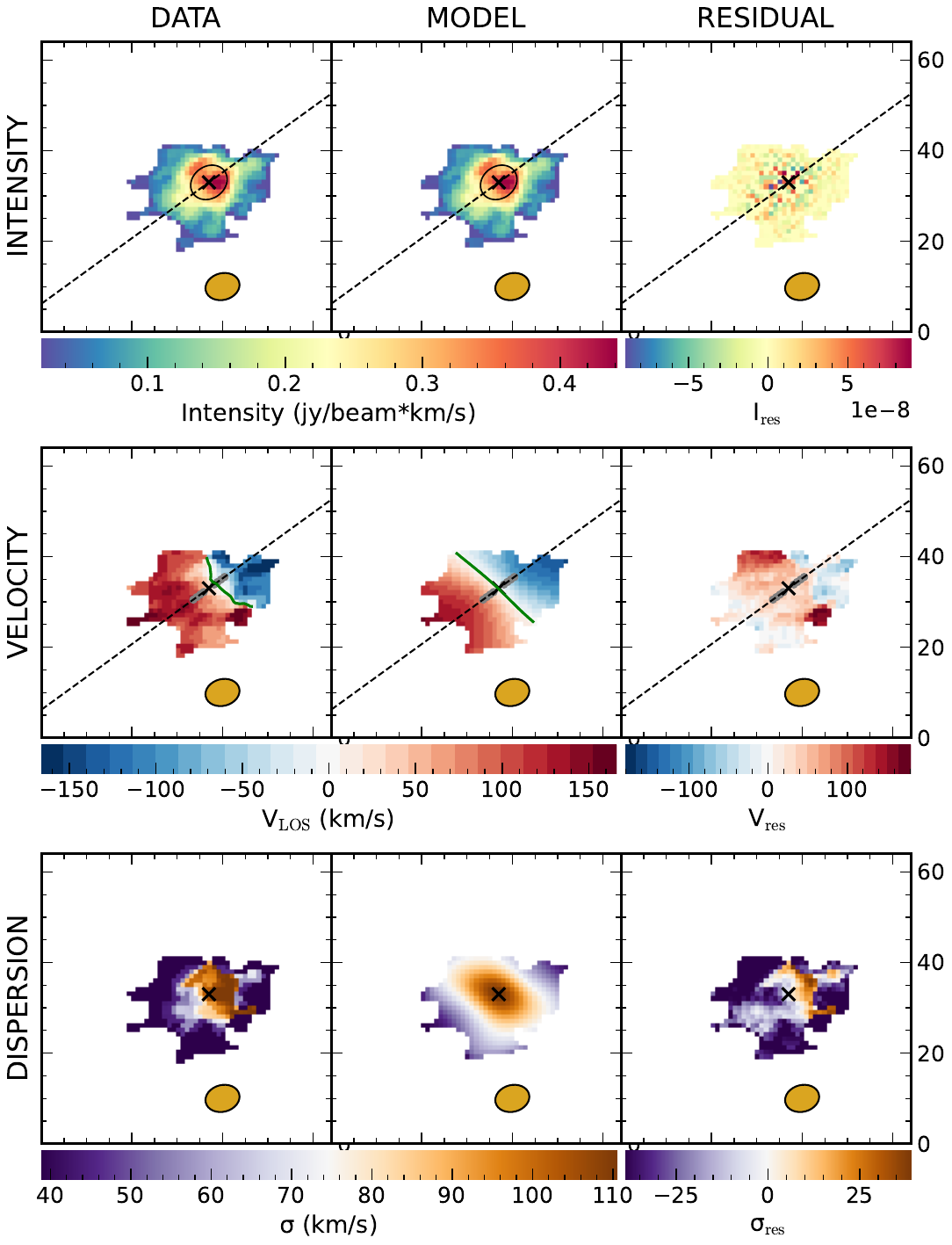}
\caption{Example fit of HAE229 using \texttt{3DBarolo}. Same as above figure.}
\label{fig:3dbarolo2}
\end{figure*}


\end{document}